\let\orilabel\label
\documentclass[reprint, amsmath, amssymb, aps]{revtex4-2}
\let\label\orilabel

\usepackage{graphicx}
\usepackage{dcolumn}
\usepackage{bm}
\usepackage{physics}
\usepackage[dvipsnames]{xcolor}
\usepackage{placeins}

\usepackage{color}
\usepackage{xcolor}
\usepackage{booktabs}
\usepackage{hyperref}

\begin{document}

\preprint{APS/123-QED}

\title{States, localization, and oxidation dynamics in gene segments with \texttt{CAG} repeats}

\author{Gerasimos Papadopoulos}
\author{Constantinos Simserides}
\email{csimseri@phys.uoa.gr}
\affiliation{Department of Physics, National and Kapodistrian University of Athens, Panepistimiopolis, Zografos GR-15784, Athens, Greece.}

\date{\today}

\begin{abstract}
We study coherent quantum-mechanical properties of gene segments with \texttt{CAG} repeats: eigenstates including density of states, mean over time probabilities of finding the hole at each site after initial oxidation at a particular site, mean transfer rates between sites, participation ratios, and inverse participation ratios. The genes are analyzed at the base-pair level within the framework of the Tight-Binding Wire Model.
All selected genes share a common feature: a region consisting of multiple repeats of the \texttt{CAG} triplet (referring to a strand in the $5'$--$3'$ direction). For each gene, there exists a range of \texttt{CAG} repeat numbers for which the gene remains functional, while values outside this range are associated with pathogenicity. We examine the genes associated with (a) Huntington’s disease, (b) Kennedy’s disease, (c) Spinocerebellar Ataxia 6, and (d) Spinocerebellar Ataxia 7.
These genes differ in the sequences flanking the \texttt{CAG} repeats, as well as in the thresholds separating non-pathogenic and pathogenic repeat lengths. In the literature, the underlying cause of these disorders is attributed to errors in DNA replication, repair and transcription leading to expansion of the triplet repeats.
We model both pathogenic and non-pathogenic sequences. We find that periodic \texttt{CAG} regions promote hole delocalization, while primers break periodicity, enhancing localization. This is of purely coherent quantum-mechanical origin, arising from the periodic structure of the \texttt{CAG} repeats and the disruption of periodicity at the primers, and cannot be explained by thermally activated hopping.
Finally, transfer depends on oxidation site, e.g., a hole generated at a \texttt{C-G} or \texttt{G-C} site is mainly transferred through other \texttt{C-G} and \texttt{G-C} base pairs, while transfer through \texttt{A-T} sites is much less probable. 
\end{abstract}

                             
\maketitle

\section{Introduction}
\label{sec:Introduction}
The aim of this work is to study charge transfer in gene segments incorporating \texttt{CAG} repeats, focusing on four relevant diseases (Huntington's disease, Kennedy's disease, Spinocerebellar Ataxias 6 and 7). Inevitably, we have to introduce the reader to biological issues, so the Introduction is about Biology.
We assume the B-DNA conformation, which is the most common structural form under physiological conditions. Genes are specific DNA sequences encoding biological information. We focus on the sequence composition at the level of base pairs, which determines the physical properties of charge movement. We use the usual symbols for the nucleic acid bases Adenine (\texttt{A}), Thymine (\texttt{T}), Guanine (\texttt{G}), Cytosine (\texttt{C}), and Uracil (\texttt{U}) in RNA.

\subsection{Genes with \texttt{CAG} repeats}
\label{subsec:CAGgenes}
\vspace{-0.3cm}
We examine four genes that share a common feature: a segment within an exon composed of repeated \texttt{CAG} trinucleotides, denoted as (\texttt{CAG})$_n$, where $n$ represents the number of repeats. The nine nucleotides immediately upstream and downstream of the (\texttt{CAG})$_n$ tract are defined as primers. 
We will realize below that primers generally stop charge delocalization. We emphasize that nine base pairs for each primer are enough to show that primers act as barriers. We arrive at the same result if we take more base pairs prior and after the \texttt{CAG} repeats as primers.
We must also clarify that in the context of our model, the ``forward primer'' and the ``reverse primer'' do not represent synthetic artifacts used in PCR (Polymerase Chain Reaction) assays. Rather, they represent the exact, endogenous, non-repetitive genomic sequences that immediately flank the \texttt{CAG} tract.
In this study, we restrict our analysis to these primers and the (\texttt{CAG})$_n$ region. For simplicity, sequences are presented in the $5' - 3'$ direction, implicitly representing the corresponding base pairing between the $5' - 3'$ and $3' - 5'$ strands. Schematically, 
\begin{equation}
\begin{array}{c}
\texttt{$5'$ ...CAG... $3'$} \\
\texttt{$3'$ ...GTC... $5'$}
\end{array} \equiv \texttt{...CAG...}
\label{eq:CAGExampleNotation}
\end{equation}
The values of $n$ that make these sequences pathogenic are denoted by $n_p$. 
Table~\ref{tab:gene_data} shows the primers and the range of $n_p$
for the four diseases studied here, specifically, 
(a) Huntington's disease, 
(b) Kennedy's disease, 
(c) Spinocerebellar Ataxia 6, 
(d) Spinocerebellar Ataxia 7. 
In Huntington’s disease $n_p =$ 36--250, located at exon 1 of the HTT gene, chromosome 4~\cite{Chintalaphani:2021},~\cite{MacDonald:1993},~\cite{McColgan:2018}.
In spinal and bulbar muscular atrophy of Kennedy (Kennedy’s disease), $n_p =$ 38--68, located at exon 1 of the AR gene, chromosome X~\cite{Chintalaphani:2021}, \cite{Fratta:2014},~\cite{Kuhlenbaumer:2001}~\cite{Spada:1991}.
In spinocerebellar ataxia 6, $n_p =$ 19--33, located at exon 47 of the CACNA1A
gene, chromosome 19~\cite{Chintalaphani:2021},~\cite{Sequeiros:2010},~\cite{Zhuchenko:1997}. 
In spinocerebellar ataxia 7, $n_p =$ 34--460, located at exon 1 of the ATXN7 gene, chromosome 3,~\cite{Chintalaphani:2021},~\cite{Cagnoli:2006},~\cite{David:1997}.

\begin{table}[htb]
\centering
\caption{Forward primer, reverse primer, and the range of pathogenic repeat expansions of the genes (where 
$n_{\mathrm{p}}$ is the number of repeats at which pathogenicity is reached).}
\footnotesize
\begin{tabular}{lccc}
\toprule
Disease & Forward primer & Reverse primer & $n_{\mathrm{p}}$ \\	\midrule
Huntington's & \texttt{AAGTCCTTC} & \texttt{CAACAGCCG} & 36--250 \\
Kennedy's    & \texttt{CTGCTGCTG} & \texttt{CAAGAGACT} & 38--68 \\
Spinocerebellar Ataxia 6 & \texttt{GGGCCCCCG} & \texttt{GCGGTGGCC} & 19--33 \\
Spinocerebellar Ataxia 7 & \texttt{GCCGCCCGG} & \texttt{CCGCCGCCT} & 34--460 \\
\bottomrule
\end{tabular}
\label{tab:gene_data}
\end{table}
\vspace{-0.2cm}

\subsection{Replication slippage in \texttt{CAG}/\texttt{CTG} repeat sequences}
\label{subsec:slippage}

The mutational instability of \texttt{CAG}/\texttt{CTG} repeats arises during replication, repair, and transcription, when DNA becomes transiently single-stranded. Under these conditions, repetitive sequences favor intra-strand misalignment, facilitating DNA polymerase slippage. This can lead to unpaired regions and, in some cases, hairpin formation. However, we focus here on the electronic structure of double-stranded DNA and do not model hairpins explicitly.

Hairpins form when complementary regions within a single strand base-pair through hydrogen bonding. In \texttt{CAG}/\texttt{CTG} repeats -- implicated in Huntington's, Kennedy's, and spinocerebellar ataxias 6 and 7 -- each strand can fold back, forming non-canonical secondary structures.

DNA polymerase copies DNA by adding complementary nucleotides in the $5' \to 3'$ direction, following \texttt{A-T} and \texttt{G-C} base-pairing rules. During replication of repetitive sequences, the polymerase may lose its position on the template, leading to misalignment. This creates a loop of extra bases, resulting in expansion (extra repeats added) or contraction (repeats skipped).

In single-stranded states, the periodicity of \texttt{CAG} repeats promotes misalignment and hairpin formation. In a \texttt{CAG} strand, \texttt{C-G} pairs stabilize the stem, while \texttt{A} residues introduce mismatches or bulges. Similarly, in the complementary \texttt{CTG} strand, \texttt{T} residues create irregularities.

\subsection{POLYQ}
\label{subsec:polyq}
\vspace{-0.3cm}
Glutamine (Gln, Q) is one of the 20 amino acids used to build proteins, encoded by the codons \texttt{CAG} and \texttt{CAA}.
Repeated \texttt{CAG} sequences in DNA lead to repeated glutamine residues in the protein, i.e., elongated polyglutamine (polyQ) tracts, associated with several neurodegenerative disorders, including those studied in this work. These disorders arise from repeat expansions in specific genes: HTT (encoding huntingtin protein), CACNA1A (a voltage-gated calcium channel subunit), ATXN7 (encoding ataxin-7), AR (encoding androgen receptor). 

\texttt{CAG} and \texttt{CAA} codons are equivalent at the protein level because they both encode glutamine (Gln). At the DNA level, they behave differently. 
\texttt{CAG} codons have the ability to form stable hairpins: They contain C and G, which can form strong C--G base pairs but also, when  single-stranded, they can fold into stable hairpins; A residues form mismatches, but the C--G stem stabilizes the structure. This creates a high propensity for polymerase slippage which creates repeat expansions.
CAA codons lack G-rich pairing opportunities and cannot form extended, stable C--G stems; they are A-rich sequences that tend to be flexible and poorly structured, i.e., they have lower tendency to form hairpins. 
Because CAA repeats form less stable secondary structures, they cause less polymerase slippage and they are less likely to expand pathologically.
Many disease genes contain mostly \texttt{CAG} repeats, interrupted by CAA codons. These interruptions act as stabilizers of the repeat tract. \texttt{CAG} triplets can form relatively stable folded states which traps the system and promotes slippage, while \texttt{CAA} triplets have fewer stabilizing interactions, the system stays more disordered and this leads to less slippage. 

Expanded polyQ tracts are toxic, forming aggregates that induce cellular stress and increase the production of reactive oxygen species (ROS), they disrupt protein homeostasis, they interfere with the protein degradation machinery, leading to the accumulation of toxic proteins. Furthermore, they contribute to loss of neuronal function. PolyQ proteins impair neurotransmitter transport and neuronal activity, ultimately causing progressive neurodegeneration and neuronal cell death, particularly in brain regions.
\texttt{CAG}/\texttt{CTG} repeats interfere with gap repair to cause repeat expansions and chromosome breaks~\cite{Polleys:2023}.
ROS interact with biomolecules such as DNA and RNA, causing oxidative damage, including the formation of 8-oxoguanine (8-oxoG). DNA repair proteins, such as OGG1, which are responsible for repairing 8-oxoG lesions, are hindered by the presence of hairpin structures. In addition, the detection of oxidative damage is impaired by hole delocalization, as discussed by Barton et al.~\cite{Barton:1996}. This finding is supported by our analysis of the participation ratio (PR) and the inverse participation ratio (IPR) of hole eigenstates in the periodic region of \texttt{CAG} repeats in genes.

\subsection{Oxidized bases and base repair proteins} 
\label{subsec:oxidizedbases}
\vspace{-0.3cm}
We briefly review selected DNA repair proteins and the chemical reactions undergone by DNA bases upon oxidation~\cite{Friedberg:2005,Dizdaroglu:2012}. 
When G is oxidized (G$^+$), a hole is formed, rendering it highly reactive. The hole can react with molecules such as water or ROS (e.g., O$_2^-$) present in the cellular environment. These reactions lead to formation of 8-oxoguanine (8-oxoG). 
Oxidation of C can lead to formation of 5-hydroxycytosine (5-OH-C).
When A is oxidized, 8-oxoadenine (8-oxoA) may form. 8-oxoA is less mutagenic than 8-oxoG, as it still preferentially pairs with T, whereas 8-oxoG can mispair with A, leading to mutations.
Oxidation of T can produce 5-hydroxymethylthymine (5-hmT).
Alternatively, thymine glycol (Tg), a more common oxidative lesion, may be generated, causing distortion of the DNA double helix. 
Several DNA repair proteins are involved in recognizing and correcting such lesions. 

Within our coherent charge transfer approach we can initially oxidize any base pair and watch the spatiotemporal evolution. This is explained in Sec.~\ref{sec:Theory}.

Theoretical descriptions of DNA charge movement have employed a variety of complementary approaches, including polaronic models  \cite{Henderson:1999, RakhmanovaConwell:2001}, atomistic Marcus-type descriptions incorporating both superexchange and hopping mechanisms \cite{CramerKrapfKoslowski:2004}, charge-transfer exciton models \cite{ConwellMcLaughlinBloch:2008,Herb:2024}, first-principles calculations of electronic couplings for hole transfer \cite{Migliore:2009}, first-principles calculations \cite{Zhang:2006}, and machine-learning prediction of DNA charge transport~\cite{KorolSegal:2019}.
While charge movement in DNA has been extensively studied experimentally and theoretically, e.g., cf. Refs.~\cite{Dandliker:1997, Rajski:2000, Genereux:2010, Zwang:2018, Wan:2000, Lewis:1997, LewisWasielewski:2013, Conron:2010, Vura-Weis:2009, Giese:2001, Giese:2006, KawaiMajima:2013, Takada:2004, FujitsukaMajima:2017}, repetitive disease-related sequences such as \texttt{CAG}/\texttt{CTG} repeats have received comparatively little attention from a charge movement perspective. We have studied charge transport (i.e., under voltage) in such cases~\cite{MLS:2023}, but not charge transfer, while it is certainly important to see what happens after e.g. an oxidation. Our previous works introduced various TB models and demonstrated that periodicity promotes extended states. Here, we extend this framework systematically to four \texttt{CAG}-repeat diseases, compare short and long repeat lengths, and analyze both static (eigenstates, DOS, PR, IPR) and dynamic properties (mean over time probabilities, mean transfer rates). We show that the periodic region supports delocalized states, while the primers act as barriers — a result that is robust across all four diseases and that provides a physical basis for understanding charge movement in these genomic regions.

In Sec.~\ref{sec:Introduction} we reviewed the relevant biological background. In Sec.~ \ref{sec:Theory} we describe how the electronic structure and coherent carrier dynamics is treated within the Tight Binding Wire Model. In Sec.~\ref{sec:ResultsDiscussion} we illustrate and discuss our results and in Sec.~\ref{sec:Conclusion} we state our conclusions.

\section{Theory}
\label{sec:Theory}  
We clarify the following key notions. By \textit{coherent} we mean quantum mechanical treatment (e.g., spatiotemporal evolution, transmission) as opposed to \textit{incoherent} which means thermally activated, semiclassical hopping.
By \textit{transfer} we mean movement without the application of an external driving force after an initial oxidation (creation of a hole) or reduction (creation of an extra electron). By \textit{transport} we mean movement under the influence of an external driving force, e.g., voltage or temperature gradient. Here we study coherent charge transfer. 

Coherent charge transport --in the same sequences related to \texttt{CAG} repeats-- has been studied in Ref.~\cite{MLS:2023}, entitled ``Charge transport properties of ideal and natural DNA segments, as mutation detectors''. Specifically, there we studied DNA sequences of ideal (textbook) geometry and of natural geometry (distorted, taken from data bases), for their coherent charge transport properties, as mutation detectors. We recruited a tight-binding (TB) wire model at the base-pair level, together with a transfer matrix technique. The relevant TB parameters were obtained using TB at the atomic level, i.e., a linear combination of all valence orbitals of all atoms. We examined quantities such as eigenspectra, density of states, transmission coefficients, and the experimentally relevant current–voltage ($I$–$V$) curves, intending to find features to recognize mutations. We introduced the normalised deviation of the $I$–$V$ curve from the origin (NDIV). NDIV seems to provide a means of evaluating the number of triplet expansions $n$; hence, we could estimate the degree of danger for developing the aforementioned diseases.

Here we use a TB wire model at the base-pair level for coherent hole transfer, with parameters extracted via another TB model at atomic level, using all valence orbitals or all atoms \cite{MSF:2021,MLS:2023}. 
Within this LCAO (linear combination of atomic orbitals) approach, the TB parameters were calculated properly, e.g., the coupling integrals (aka interaction or transport or transfer or hopping integrals) between base pairs are affected when we change the relative orientation of base pairs \texttt{X-Y} and \texttt{Z-W}, i.e.,
\begin{equation}
\begin{array}{c}
\texttt{$5'$ ...X-Y... $3'$} \\
\texttt{$3'$ ...Z-W... $5'$}
\end{array} 
\texttt{ to }
\begin{array}{c}
\texttt{$5'$ ...X-Y... $3'$} \\
\texttt{$3'$ ...W-Z... $5'$}
\end{array} 
\label{eq:XYZWtoXYWZ}
\end{equation}
In Ref.~\cite{MSF:2021} we compared our LCAO method with experimental and ab initio ionization and excitation energies of many purines and pyrimidines. So, the TB wire model that we use here is not a naive choice; it is a choice that focuses on the codon (the triplet of base pairs) \texttt{CAG} $\equiv$ \texttt{CTG}, i.e. on 
\begin{equation}
\begin{array}{ccccccccc}
5' & ... & \texttt{C} & \texttt{A} & \texttt{G} & ... & 3' \\
& & \texttt{|} & \texttt{|} & \texttt{|} & & \\
3' & ... & \texttt{G} & \texttt{T} & \texttt{C} & ... & 5'
\end{array}
\label{eq:CAGGTC}
\end{equation}
We have used various other coarse-grained TB models in the past, e.g., the extended ladder model~\cite{LChMKLThTS:2016, LKMTLGThChS:2016}, the four-channel model~\cite{KMS:2025}, the fishbone wire model~\cite{BFS:2025}, many models and coupling to the environment~\cite{HTMSAR:2025}. For example, an (extended) ladder model is more detailed than the TB wire model, of course. Detailed comparisons of the TB wire with the TB (extended) ladder model can be found in Refs.~\cite{LChMKLThTS:2016, LKMTLGThChS:2016}. Within our present TB wire model, the HOMO on-site energy of \texttt{G-C}, $E_\texttt{G-C} = -8.30$ eV and the HOMO on-site energy of \texttt{A-T}, $E_\texttt{A-T} = -8.49$ eV. The absolute values of the interaction (or coupling) integrals are listed in Table~\ref{Table:InteractionIntegrals}. 

\begin{table}[hb]
\centering
\begin{tabular}{cc|cc|cc}
\toprule
\multicolumn{6}{c}{dimer and $|t|$ (eV)} \\
\midrule
\texttt{AA} $\equiv$ \texttt{TT} & 0.038 & \texttt{GG} $\equiv$ \texttt{CC} & 0.116 & \texttt{AG} $\equiv$ \texttt{CT} & 0.037 \\
\texttt{TA}             & 0.037 & \texttt{CG}             & 0.075 & \texttt{GA} $\equiv$ \texttt{TC} & 0.142 \\
\texttt{AT}             & 0.050 & \texttt{GC}             & 0.010 & \texttt{GT} $\equiv$ \texttt{AC} & 0.016 \\
               &       &                &       & \texttt{TG} $\equiv$ \texttt{CA} & 0.028 \\
\bottomrule
\end{tabular}
\caption{Absolute values of the interaction integrals, $|t|$, for the HOMO regime (holes). By \texttt{XY} we mean the $5'$ \texttt{XY} $3' \equiv 5'$ \texttt{Y}$_\mathrm{c}$\texttt{X}$_\mathrm{c}$ $3'$ dimer (c stands for ``complementary''). }
\label{Table:InteractionIntegrals}
\end{table}

In a TB wire model, \textit{site} or \textit{monomer} or \textit{base pair} are equivalent terms. The physical quantities that we study have been presented, e.g., in Refs.~\cite{LChMKLThTS:2016}, \cite{Simserides:2014},  \cite{LChMKThS:2015}, \cite{LVBMS:2018}. 
We call $\mu$ the monomer index, $\mu = 1,2,\dots,N$. We just mention the $5'-3'$ base sequence of one strand, e.g., we denote two successive monomers by \texttt{YX}, meaning that the base pair \texttt{X-X}$_{\text{c}}$ is separated and twisted by 3.4 {\AA} and $36^{\circ}$, respectively, relatively to the base pair \texttt{Y-Y}$_{\text{c}}$, around the B-DNA growth axis. \texttt{X}$_{\text{c}}$ (\texttt{Y}$_{\text{c}}$) is the complementary base of \texttt{X} (\texttt{Y}).  We assume that the state or movement of an extra hole or electron can be expressed through the monomer HOMOs or LUMOs, respectively, cf. Eqs.~\eqref{TIP} and \eqref{TDP} below. We will show results for oxidation, i.e., relative to the creation of a hole. 

For the reader that might be interested, what happens at the base level ~\cite{LChMKLThTS:2016, LKMTLGThChS:2016} is the following. Suppose we have the sequence:
\begin{equation}
\begin{array}{ccccccccccccc}
5' & ... & \texttt{C} & \texttt{A} & \texttt{G} & \texttt{C} & \texttt{A} & \texttt{G} & ... & 3' \\
   &     & \texttt{|} & \texttt{|} & \texttt{|} & \texttt{|} & \texttt{|} & \texttt{|} &     & \\
3' & ... & \texttt{G} & \texttt{T} & \texttt{C} & \texttt{G} & \texttt{T} & \texttt{C} & ... & 5'
\end{array}
\label{eq:CAGCAGGTCGTC}
\end{equation}
When we oxidize a \texttt{C-G}  (\texttt{C} on $5' - 3'$ strand, \texttt{G} on $3' - 5'$ strand), the hole will be mainly created in \texttt{G} due to the lower ionization energy of \texttt{G} compared to \texttt{C}. In other words, the HOMO of \texttt{C-G} is very close to the HOMO of \texttt{G}. Then, the hole will travel mainly through \texttt{G}s which are diagonally located. 
When we oxidize a \texttt{G-C} (\texttt{G} on $5' - 3'$ strand, \texttt{C} on $3' - 5'$ strand), the hole will be mainly created in \texttt{G} due to the lower ionization energy of \texttt{G} compared to \texttt{C}. In other words, the HOMO of \texttt{G-C} is very close to the HOMO of \texttt{G}. Then, the hole will travel mainly through \texttt{G}s which are diagonally located. 
When we oxidize an \texttt{A-T} (\texttt{A} on $5' - 3'$ strand, \texttt{T} on $3' - 5'$ strand), the hole will be mainly created in \texttt{A} due to the lower ionization energy of \texttt{A} compared to \texttt{T}. In other words, the HOMO of \texttt{A-T} is very close to the HOMO of \texttt{A}. Then, the hole will travel mainly through \texttt{A}s which are diagonally located. 
Interestingly, if we oxidize a \texttt{C} on purpose, the hole will mainly move through \texttt{C}s because in this coherent regime sites with identical on-site energies are favored. Similarly in the case of oxidizing on purpose a \texttt{T}. These happen at the base level \cite{LChMKLThTS:2016, LKMTLGThChS:2016}. However, here we study the most probable oxidations.  
The loss of transverse (inter-strand) resolution does not invalidate the finding that charge localizes differently in pure periodic tracts versus disordered flanking regions. The wire model provides a robust approximation that successfully captures the macroscopic order-to-disorder transitions characterizing the pathogenic expansion boundaries. 
We think that it is not necessary to use one more TB coarse-grained model, like the (extended) ladder model in this work.

\subsection{Time-independent problem} 
\label{subsec:TIproblem}
\vspace{-0.3cm}
The TB \textit{wire} model Hamiltonian can be written as
\begin{equation} \label{WireHamiltonian}
\hat{H}_\text{W} =
\sum_{\mu=1}^{N} E_\mu \ketbra{\mu}{\mu} +
\bigg( \sum_{\mu=1}^{N-1} t_{\mu,\mu+1} \ketbra{\mu}{\mu+1} + h.c. \bigg).
\end{equation}
$E_\mu$ is the on-site energy of the $\mu$-th site, and $t_{\mu,\lambda} = t_{\lambda,\mu}^*$ is the interaction integral between sites $\mu$ and $\lambda$. 
The state of a polymer can be expressed as
\begin{equation} \label{TIP}
\ket{\mathcal{P}} = \sum_{\mu=1}^{N} v_\mu \ket{\mu}.
\end{equation}
Substituting Eqs. \eqref{WireHamiltonian} and \eqref{TIP} to the time-independent Schr\"odinger equation
\begin{equation} \label{TISchr}
\hat{H} \ket{\mathcal{P}} = E \ket{\mathcal{P}},
\end{equation}
we arrive to a system of $N$ coupled equations
\begin{equation} \label{TIsystembp}
E_\mu v_{\mu} + t_{\mu, \mu+1} v_{\mu+1} + t_{\mu, \mu-1} v_{\mu-1} = E v_{\mu}, 
\end{equation}
which is equivalent  to the eigenvalue-eigenvector problem
\begin{equation} \label{eigenvproblem}
H\vec{v} = E \vec{v},
\end{equation}
where $H$ is the hamiltonian matrix of order $N$, composed of the TB parameters $E_\mu$ and $t_{\mu,\lambda}$, and $\vec{v}$ is the vector matrix composed of the coefficients $v_\mu$. The diagonalization of $H$ leads to the determination of the eigenenergy spectrum (\textit{eigenspectrum}), $\{E_k\}$, $k =1,2,\dots,N$, for which we suppose that $E_1<E_2<\dots<E_{N}$, as well as to the determination of the occupation probabilities for each eigenstate, $\abs{v_{\mu k}}^2$, where $v_{\mu k}$ is the $\mu$-th component of the $k$-th eigenvector. $\{v_{\mu k}\}$ are normalized, and their linear independence is checked in all cases.

Having determined the eigenspectrum, the density of states (DOS), generally defined by
\begin{equation} \label{DOS}
g(E) = \sum_{k=1}^{N} \delta(E-E_k),
\end{equation}
can be computed. Changing the view of a polymer from one (e.g. top) to the other (e.g. bottom) side of the growth axis, reflects the hamiltonian matrix $H$ of the polymer on its main antidiagonal. This reflected Hamiltonian, $H^{\text{eq}}$, describes the \textit{equivalent polymer}~\cite{LChMKLThTS:2016}. 
$H$ and $H^{\text{eq}}$ are connected by the similarity transformation $H^{\text{eq}} = J^{-1}HJ$, where $J(=J^{-1})$ is the unit antidiagonal matrix of order $N$.
Therefore, $H$ and $H^{\text{eq}}$ have identical eigenspectra (hence the equivalent polymers' DOS is identical) and their eigenvectors are connected by $v_{\mu k} = v_{(N-\mu+1)k}^{\text{eq}}$. Generally,
$\text{eq(\text{XY\dots Z})} =
\text{Z}_{\text{c}}\dots\text{Y}_{\text{c}}\text{X}_{\text{c}}.$

The participation ratio (PR) \cite{Wegner:1980} quantifies the localization of eigenstates, i.e., the extent an eigenstate is localized on one or more sites, in our case, base pairs or monomers. For a system with $N$ sites, the participation ratio of eigenstate $k$ is defined as
\begin{equation}
\label{eq:PR}
\mathrm{PR}_{k} = \frac{1}{N \sum_{\alpha=1}^{N} |v_{\alpha k}|^{4}},
\end{equation}
where $v_{\alpha k}$ is the site-$\alpha$ component of eigenvector $k$, corresponding to the eigenenergy $E_k$. The eigenvectors are normalized, i.e, $\sum_{\alpha=1}^{N} |v_{\alpha k}|^{2}=1, \forall k$.
$\mathrm{PR}_k \approx 1/N$ indicates strong localization, i.e., the state is concentrated approximately on one site, while
$\mathrm{PR}_k \approx 1$ indicates uniform delocalization, i.e., the state is evenly spread over all $N$ sites.
For \textit{degenerate} eigenstates, the standard definition becomes ambiguous because any linear combination of degenerate eigenvectors is an eigenvector. When eigenstates $\{k_1, k_2, \dots, k_d\}$ form a $d$-fold degenerate subspace ($E_{k_1} = E_{k_2} = \cdots = E_{k_d}$), the proper definition is
\begin{equation}
\label{eq:PR_degenerate}
\mathrm{PR}_{k_j} = \frac{d^2}{N \sum_{\alpha=1}^{N} \left( \sum_{j=1}^{d} |v_{\alpha k_j}|^{2} \right)^{2}},
\end{equation}
for $k_j \in$ degenerate subspace ($j=1, \dots, d$).
This expression is invariant under unitary rotations within the degenerate subspace and measures the collective spreading of the entire eigenspace. For all eigenstates $k_1, \dots k_d$ within the same degenerate subspace, the generalized participation ratio,  Eq.~\eqref{eq:PR_degenerate}, assigns them the same PR value. For $d=1$, it reduces to the standard definition, Eq.~\eqref{eq:PR}. The Inverse Participation Ratio (IPR) is also often discussed. The relation between PR and IPR is
\begin{equation}
\mathrm{IPR}_k = \sum_{\alpha=1}^{N} |v_{\alpha k}|^{4}, \quad
\mathrm{PR}_k = 1/(N \times \mathrm{IPR}_k).
\end{equation}
The $q$-th moment or generalized IPR \cite{Mirlin:2000, Cuevas:2001, Evers:2008} is
\begin{equation}
\text{IPR}_q = \sum_{\alpha=1}^{N} |v_{\alpha k}|^{2q}.
\end{equation}
$ q=2 $ gives the standard IPR. We underscore some special cases: 
For a non-degenerate state localized at one site, $\mathrm{PR}_k = 1/N$. For a non-degenerate fully extended state, $\mathrm{PR}_k = 1$.
For $d$-fold degenerate localized states (each localized on different sites), $\mathrm{PR}_{\mathrm{subspace}} = d/N$.
For $d$-fold degenerate extended states (uniformly spread), $\mathrm{PR}_{\mathrm{subspace}} = 1$.
The PR thus provides a unified measure of localization for both non-degenerate and degenerate eigenstates, with values normalized to the range $[1/N, 1]$.

\subsection{ Time-dependent problem}
\label{subsec:TDproblem} 
\vspace{-0.3cm}
To describe the spatiotemporal evolution of an extra carrier (hole/electron), inserted or created (e.g. by oxidation/reduction) in a particular monomer of the polymer, we consider the state of the polymer as
\begin{equation} \label{TDP}
\ket{\mathcal{P}(t)} = \sum_{\mu=1}^{N}C_\mu(t) \ket{\mu},
\end{equation}
where $\abs{C_\mu(t)}^2$ is the probability to find the carrier at the $\mu$-th monomer at time $t$. Substituting Eqs. \eqref{WireHamiltonian} and \eqref{TDP} to the time-dependent Schr\"odinger equation
\begin{equation} \label{TDSchr}
i\hbar \pdv{t}\ket{\mathcal{P}(t)} = \hat{H} \ket{\mathcal{P}(t)},
\end{equation}
we arrive at a system of $N$ coupled differential equations 
\begin{equation} \label{TDsystembp}
i\hbar \dv{C_\mu}{t} = E_\mu C_\mu + t_{\mu, \mu+1} C_{\mu+1} + t_{\mu, \mu-1} C_{\mu-1}.
\end{equation}
Eq. \eqref{TDsystembp} is equivalent  to a 1$^\text{st}$ order matrix differential equation of the form
\begin{equation} \label{Matrixdiffeq}
\dot{\vec{C}}(t) = -\frac{i}{\hbar}H \vec{C}(t),
\end{equation}
where $\vec{C}(t)$ is a vector matrix composed of the coefficients $C_\mu(t), \;\; \mu = 1, 2, \dots, N$. Since $\vec{C}(t)$ has both vector and time-dependent character, Eq.~\eqref{Matrixdiffeq} can be solved by looking for solutions of the form $\vec{C}(t) = \vec{v} e^{-\frac{i}{\hbar}Et}$, separating the vector character and the time-dependent character. Then, $\dot{\vec{C}}(t) = -\frac{i}{\hbar}E \vec{v}e^{-\frac{i}{\hbar}Et}$. Substituting,
$\vec{C}(t)$ and $\dot{\vec{C}}(t)$, Eq.~\eqref{Matrixdiffeq} leads to the eigenvalue problem of Eq.~\eqref{eigenvproblem}, that is, $H\vec{v} = E \vec{v}$. Having determined the eigenvalues $E_k$ and eigenvectors $\vec{v}_k$ of $H$, each $\vec{v}_k e^{-\frac{i}{\hbar}E_kt}$ is a partial solution of Eq.~\eqref{Matrixdiffeq}. Hence, the general solution of Eq.~\eqref{Matrixdiffeq} is
\begin{equation} \label{generalsolution}
\vec{C}(t) = \sum_{k=1}^{N} \sigma_k \vec{v}_k e^{-\frac{i}{\hbar}E_kt},
\end{equation}
where $\sigma_k$ are determined from the initial conditions: If we define the $N \times N$ eigenvector matrix $V$, with elements $v_{\mu k}$ [$\mu$-th component (relevant to monomer $\mu$) of the eigenvector $k$], the vector matrix $\vec{\Sigma}$, composed of the coefficients $\sigma_k, \;\; k = 1, 2, \dots, N$, is given by
\begin{equation} \label{sigma_kdef}
\vec{\Sigma} = V^T\vec{C}(0).
\end{equation}
Suppose that initially the extra carrier is placed at the $\lambda$-th monomer, i.e., $C_\lambda(0) = 1$, $C_\mu(0) = 0, \forall \mu \neq \lambda$. Then,
\begin{equation} \label{Sigmamatrix}
\vec{\Sigma} = \begin{bmatrix}
v_{\lambda 1}\\\vdots\\v_{\lambda k}\\\vdots\\v_{\lambda N}
\end{bmatrix}.
\end{equation}
In other words, the coefficients $\sigma_k$ are given by the row of the eigenvector matrix which corresponds to the monomer the carrier is initially placed at.

From Eq.~\eqref{generalsolution} it follows that the probability to find the carrier at the $\mu$-th monomer is
\begin{equation} \label{probabilities}
\abs{C_\mu(t)}^2 = \sum_{k=1}^{N} \sigma_k^2v_{\mu k}^2 + 2 \sum_{k=1}^{N} \sum_{\substack{k'=1 \\k'<k}}^{N} \sigma_k \sigma_{k'} v_{\mu k} v_{\mu k'} \cos(2\pi f_{kk'}t),
\end{equation}
\begin{equation} \label{fandT}
f_{kk'} = \frac{1}{T_{kk'}} = \frac{E_k-E_{k'}}{h}, \; \forall k > k',
\end{equation}
are the frequencies ($f_{kk'}$) or periods ($T_{kk'}$) involved in charge transfer. If $m$ is the number of discrete eigenenergies, then, the number of different $f_{kk'}$ or $T_{kk'}$ involved in carrier transfer is $S= \binom{m}{2} = \frac{m!}{2!(m-2)!} = \frac{m(m-1)}{2}$.
If there are no degenerate eigenenergies (which holds for all cases studied here, but e.g. does not hold for \textit{cyclic} I1 polymers~\cite{LChMKThS:2015}), then $m = N$. If eigenenergies are symmetric relative to some central value, then, $S$ decreases (there exist degenerate $f_{kk'}$ or $T_{kk'}$). Specifically, in that case, $S = \frac{m^2}{4}$, for even $m$ and $S = \frac{m^2-1}{4}$ for odd $m$.

The frequency content of the probability oscillations can be analyzed as follows: From Eq.~\eqref{probabilities} it can be shown that the one-sided Fourier amplitude spectrum that corresponds to the probability $\abs{C_\mu(t)}^2$ is given by
\begin{equation} \label{Fourierspectra}
\resizebox{\hsize}{!}{$\displaystyle
\abs{\mathcal{F}_\mu(f)} = \sum_{k=1}^{N} \sigma_k^2 v_{\mu k}^2 \delta(f) + 2 \sum_{k=1}^{N} \sum_{\substack{k'=1 \\ k'<k}}^{N} \abs{\sigma_k \sigma_{k'} v_{\mu k} v_{\mu k'}} \delta(f-f_{kk'})$},
\end{equation}
where $2 |\sigma_k v_{\mu k} \sigma_{k'} v_{\mu k'}|$ is the Fourier amplitude of frequency $f_{kk'}$.
The Weighted Mean Frequency (WMF) of monomer $\mu$ can be defined by
\begin{equation} \label{Eq:WMFdef}
f_{WM}^\mu = \frac{\displaystyle\sum_{k=1}^{N} \sum_{\substack{k'=1 \\k'<k}}^{N} |\sigma_k v_{\mu k} \sigma_{k'} v_{\mu k'}| f_{kk'}}{
\displaystyle \sum_{k=1}^{N} \sum_{\substack{k'=1 \\k'<k}}^{N} |\sigma_k v_{\mu k} \sigma_{k'} v_{\mu k'}|}.
\end{equation}
WMF expresses the mean frequency content of the  carrier oscillation at monomer $\mu$. Having determined the WMF for all monomers, we can now obtain a measure of the overall frequency content of carrier oscillations in the polymer: Since $f_{WM}^\mu$ is the weighted mean frequency of monomer $\mu$ and $\left\langle|{C_\mu(t)|^2}\right\rangle$ is the mean over time probability of finding the extra carrier at monomer $\mu$, we define the Total Weighted Mean Frequency (TWMF) as
\begin{equation} \label{Eq:TWMFdef}
f_{TWM} = \sum_{\mu=1}^{N} f_{WM}^\mu \left\langle|C_\mu(t)|^2\right\rangle.
\end{equation}
Similar analysis can be performed for the electric dipole moment oscillations. 

From Eq.~\eqref{probabilities} it follows that the mean over time probability to find the extra carrier at the $\mu$-th monomer is
\begin{equation} \label{meanprobabilities}
\expval{\abs{C_\mu(t)}^2} = \sum_{k=1}^{N} \sigma_k^2v_{\mu k}^2,
\end{equation}
provided there are no degenerate eigenstates.
In the case of initial excitation only at site 
$\lambda$, using Eq.~\eqref{Sigmamatrix}, we get
\begin{equation}
\langle |C_{\lambda \mu }(t)|^2 \rangle =
\sum_{k=1}^N 
v_{\lambda k}^2
v_{\mu k}^2 
\end{equation}
where we have added the index $\lambda$ indicating the site of initial excitation.
Also, under the previous assumptions, it follows that 
$
\langle |C_{\lambda \mu}(t)|^2 \rangle = 
\langle |C_{\mu \lambda}(t)|^2 \rangle$.

A quantity that evaluates simultaneously the magnitude of charge transfer and the time scale of the phenomenon, is the mean transfer rate~\cite{Simserides:2014}
\begin{equation} \label{pmtr}
k_{\lambda \mu} = \frac{\expval{\abs{C_\mu(t)}^2}}{t_{\lambda \mu}}.
\end{equation}
$t_{\lambda \mu}$ is the \textit{mean transfer time}, i.e.,
having placed the carrier initially at monomer $\lambda$,
the time it takes for the probability to find the extra carrier at monomer $\mu$,
$\abs{C_\mu(t)}^2$, to become equal to its mean value, $\expval{\abs{C_\mu(t)}^2}$, for the first time. 
For the mean transfer rates it holds
$ k_{\lambda \mu} = k_{\mu \lambda}=k_{(N-\lambda+1)(N-\mu+1)}^{\text{equiv}} = k_{(N-\mu+1)(N-\lambda+1)}^{\text{equiv}}$.

\section{Results and Discussion}
\label{sec:ResultsDiscussion}
We present and discuss our results for the previously introduced physical quantities, supported by the corresponding plots, for the four genes under investigation. For each gene, we consider an odd number of repeats $n$ of the \texttt{CAG} sequence, ensuring the presence of a central triplet. Flanking the \texttt{(CAG)}$_n$ segment are non-periodic regions, namely the forward and reverse primers, each of length 9 base pairs.
As we will show below, nine base pairs for each primer are sufficient to demonstrate that the primers inhibit charge delocalization. The same conclusion is reached when longer primer sequences are considered on both sides of the \texttt{CAG} repeats.
In our indexing, monomer 1 corresponds to the $5'$ end of the forward primer, and monomer $N$ corresponds to the $3'$ end of the reverse primer.

We employ two different values of $ n $: a small value ($ n = 3 $) and a larger one within the pathological range. The results for $n=3$ are shown in  Appendix~\ref{sec:n3plots}.
The values $n_p = 37, 39, 25, 37$ that we use for Huntington's disease, Kennedy's disease, Spinocerebellar Ataxia type 6, Spinocerebellar Ataxia type 7, respectively, were selected as representative examples within the (lower) corresponding pathological repeat ranges reported in the literature.
Increasing the number of repeats further just makes the diagrams more difficult to discern. 
They were not intended to represent specific patients. Since our objective is to investigate the effect of the presence of a pathogenic repeat expansion rather than patient-specific repeat lengths, the qualitative conclusions of our study are not expected to depend on the exact pathological value chosen.
The smaller auxiliary (reference) sequence is used to facilitate clearer visualization of certain features. This auxiliary sequence contains $ n = 3 $ \texttt{CAG} repeats in the central region (9 base pairs); including the flanking primer regions, this yields a total length of 27 base pairs.

The plots are organized into two groups: the first corresponds to time-independent quantities characterizing the electronic structure; the second concerns time-dependent quantities characterizing charge transfer. For this reason, Subsec.\ref{subsec:TI} is devoted to time-independent quantities and  Subsec.\ref{subsec:TD} is devoted to time-dependent quantities.
The density of states (DOS) per number of monomers is based on counting the number of eigenstates per energy window. In the DOS figures we use 20000 windows to show as much detail as possible. We have checked that the DOS calculated numerically coincides with known analytical results~\cite{LS:2018} for $N \to \infty$.
For the time-dependent analysis, we introduce an oxidation event at a base pair within the central \texttt{CAG} triplet and examine the system’s response. Table~\ref{table:repeats} summarizes the values of the repeat number $ n $ and the total number of base pairs $ N $ in the sequences studied. Since two primers of 9 base pairs each are included, the total number of monomers, $ N = 3n + 18 $.

\begin{table}[!t]
\centering
\caption{Gene characteristics: number of \texttt{CAG} repeats ($n$) in the central triplet region and total number of base pairs ($N$) in the sequence. The total number of monomers, $N = 3n + 18$.
Results for the auxiliary sequences are shown in the Appendix.}
\begin{tabular}{lcc}
\toprule
Disease & \texttt{CAG} repeats ($n$) & Base pairs ($N$) \\
\midrule
Huntington's & 37 & 129 \\
Kennedy's & 39 & 135 \\
Spinocerebellar Ataxia 6 & 25 & 93 \\
Spinocerebellar Ataxia 7 & 37 & 129 \\
Auxiliary sequences & 3 & 27 \\
\bottomrule
\end{tabular}
\label{table:repeats}
\end{table}
\vspace{-0.2cm}

\subsection{Time-independent}
\label{subsec:TI}
\begin{figure}[h]
\centering
\hspace{-0.5cm}
\includegraphics[width=0.24\textwidth]{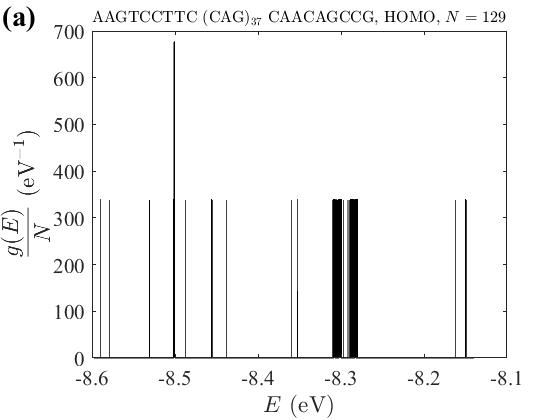}
\includegraphics[width=0.24\textwidth]{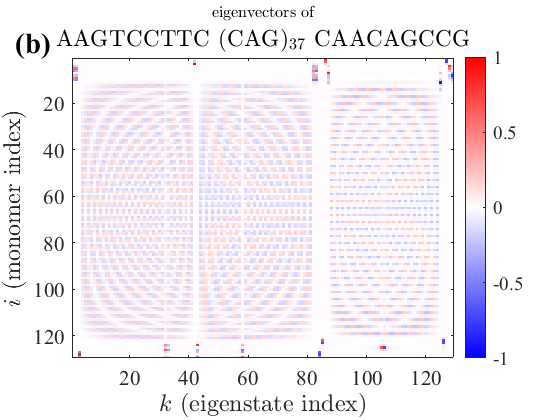}\\
\hspace{-0.5cm}
\includegraphics[width=0.24\textwidth]{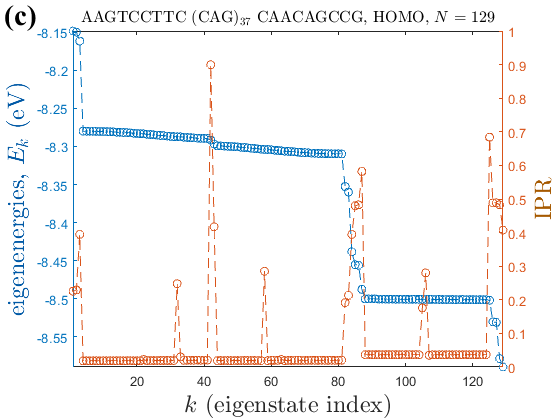}
\includegraphics[width=0.24\textwidth]{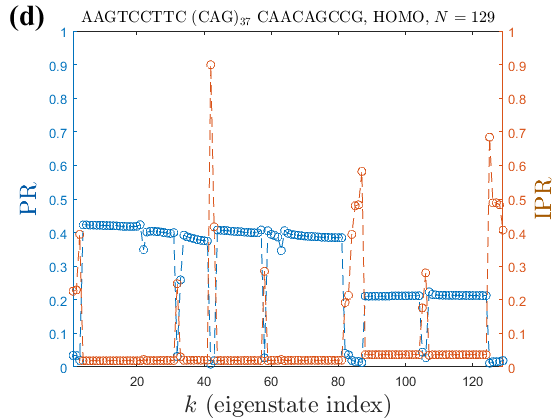}
\caption{Huntington’s disease, for  $n=37$ ($N=129$): 
(a) density of States (DOS), (b) eigenvectors, (c) eigenenergies $E_k$ with inverse participation ratio (IPR), and (d) participation ratio (PR) with IPR.}
\label{fig:HuntingtonAllTIn37}
\end{figure}

The main features of the time-independent problem for Huntington’s disease are illustrated in Fig.~\ref{fig:HuntingtonAllTIn37} ($n=37$, $N=129$) and Fig.~\ref{fig:HuntingtonAllTIn3} ($n=3$, $N=27$). Each figure has four panels: (a) density of states (DOS), (b) eigenvectors, (c) eigenenergies $E_k$ with inverse participation ratio (IPR), (d) participation ratio (PR) with IPR.
IPR appears in panel (c) together with eigenspectra because this way we display clearly the three subbands with relatively delocalized states and the few very localized states between them. 
PR and IPR are both shown in panel (d) because PR emphasizes delocalized states while IPR emphasizes localized states. These four panels show different aspects of how the periodic \texttt{CAG} region promotes extended states, whereas primers introduce disruption of periodicity enhancing localization at the edges of the sequence. Moreover, increasing $n$ enhances periodicity, leading to smoother spectra, higher PR values, and a more structured DOS: the reader may compare Fig.~\ref{fig:HuntingtonAllTIn37} ($n=37$, $N=129$) with 
Fig.~\ref{fig:HuntingtonAllTIn3} ($n=3$, $N=27$) in the Appendix. We notice that many states are relatively delocalized but some states are strongly localized; for example, for $n=3$, the $k=8$ eigenstate is localized at the 3rd monomer (a \texttt{G-C} base pair of the forward primer), while the $k=21$ eigenstate is localized at the 23rd monomer (an \texttt{A-T} base pair of the reverse primer). Other eigenstates form the \texttt{G-C} and \texttt{C-G} subbands (close to the \texttt{G-C} on-site energy), as well as the \texttt{A-T} subband (close to the \texttt{A-T} on-site energy).

\begin{figure}[h]
\centering
\hspace{-0.5cm}
\includegraphics[width=0.24\textwidth]{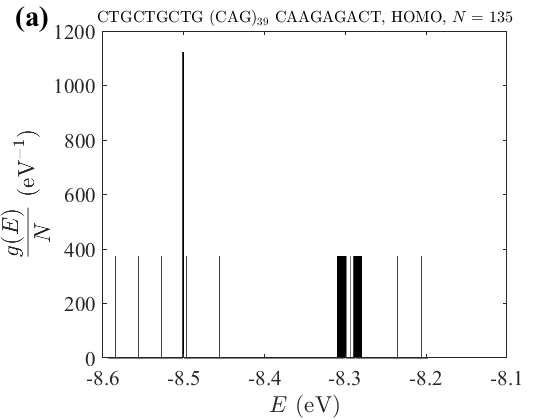}
\includegraphics[width=0.24\textwidth]{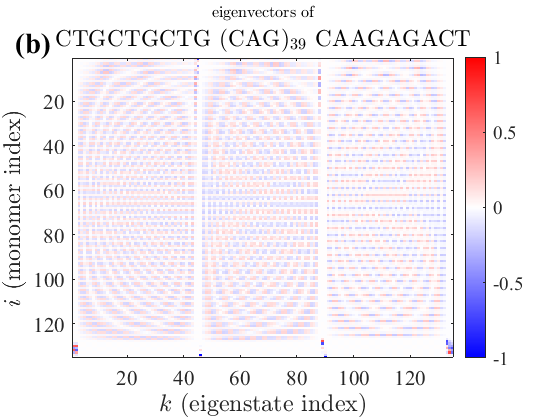}\\
\hspace{-0.5cm}
\includegraphics[width=0.24\textwidth]{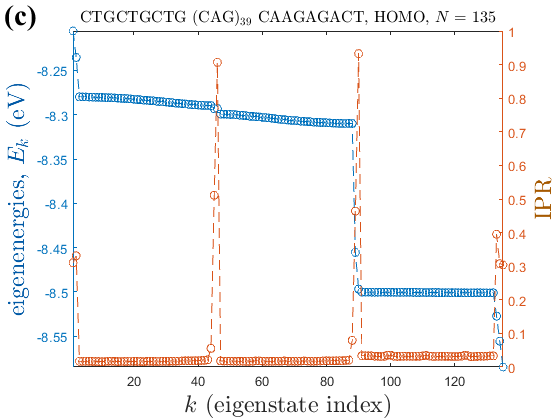}
\includegraphics[width=0.24\textwidth]{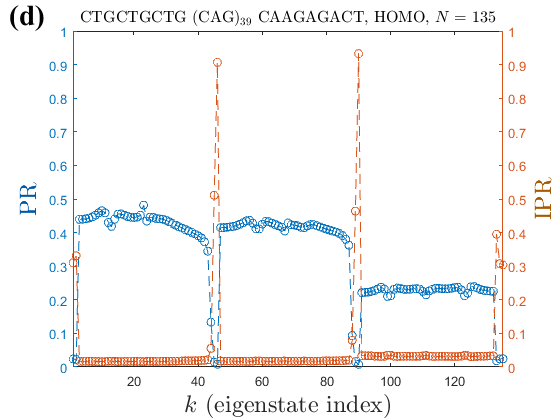}
\caption{Kennedy’s disease, for $n=39$ ($N=135$): 
(a) density of States (DOS), (b) eigenvectors, (c) eigenenergies $E_k$ with inverse participation ratio (IPR), and (d) participation ratio (PR) with IPR.}
\label{fig:KennedyAllTIn39}
\end{figure}

For Kennedy’s disease, the time-independent features are shown in Fig.~\ref{fig:KennedyAllTIn39} ($n=39$, $N=135$) and in Fig.~\ref{fig:KennedyAllTIn3} ($n=3$, $N=27$). We observe behavior similar to that of Huntington’s disease (cf. Fig.~\ref{fig:HuntingtonAllTIn37} and  Fig.~\ref{fig:HuntingtonAllTIn3}), with one difference: the forward primer closely resembles the periodic segment, differing only by the inversion of base pair \texttt{A-T} to \texttt{T-A}. This leads to a smoother transition from the central periodic region to the forward primer, since the on-site energies are identical. In contrast, the reverse primer inhibits delocalization. The density of states is similar to that of Huntington’s disease for large values of $n$.
We further observe that eigenstates with significant contribution from the forward primer do not differ substantially from those of the periodic segment. This is expected, as the same sequence is present, but with reversed \texttt{A-T} base pairs. The periodicity of the \texttt{CAG} repeats favors higher PR values, indicating a stronger contribution to extended states. In contrast, states with strong contribution from the reverse primer exhibit noticeable disorder: the reverse primer induces localization, thereby hindering hole transfer.

\begin{figure}[h]
\centering
\hspace{-0.5cm}
\includegraphics[width=0.24\textwidth]{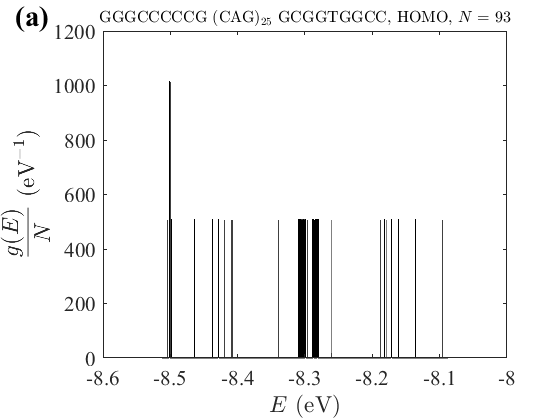}
\includegraphics[width=0.24\textwidth]{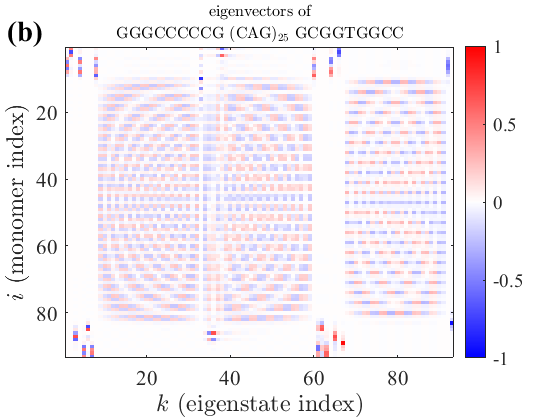}\\
\hspace{-0.5cm}
\includegraphics[width=0.24\textwidth]{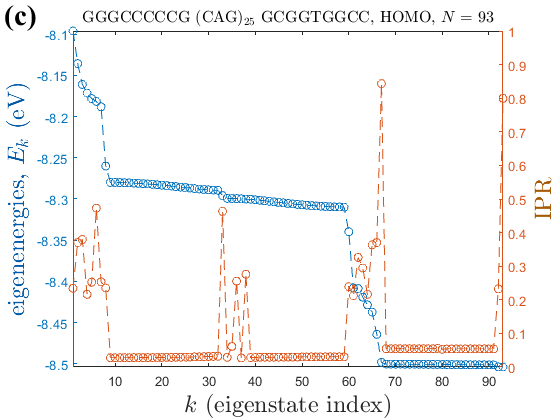}
\includegraphics[width=0.24\textwidth]{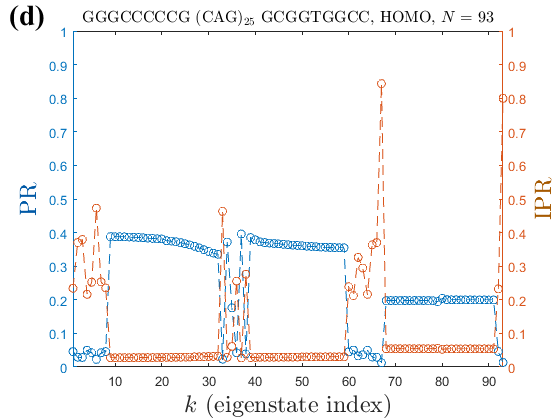}
\caption{Spinocerebellar Ataxia 6, for  $n=25$ ($N=93$): 
(a) density of States (DOS), (b) eigenvectors, (c) eigenenergies $E_k$ with inverse participation ratio (IPR), and (d) participation ratio (PR) with IPR.}
\label{fig:Ataxia6AllTIn25}
\end{figure}

For Spinocerebellar Ataxia 6, the time-independent features are illustrated in Fig.~\ref{fig:Ataxia6AllTIn25} ($n=25$, $N=93$) and in Fig.~\ref{fig:Ataxia6AllTIn3} ($n=3$, $N=27$). Again, the periodicity of the \texttt{CAG} repeats in the central region promotes extended states, which is important for charge movement. The primers introduce disorder, hindering the extension of states toward the edges. This is pronounced at the reverse primer, as the forward primer only contains \texttt{G-C} or \texttt{C-G} base pairs and acts as an almost periodic part but with different periodicity relative to the (\texttt{CAG})$_n$ central segment. Increasing the number of repeats $n$, enhances periodicity, thereby positively affecting charge movement.

\begin{figure}[h]
\centering
\hspace{-0.5cm}
\includegraphics[width=0.24\textwidth]{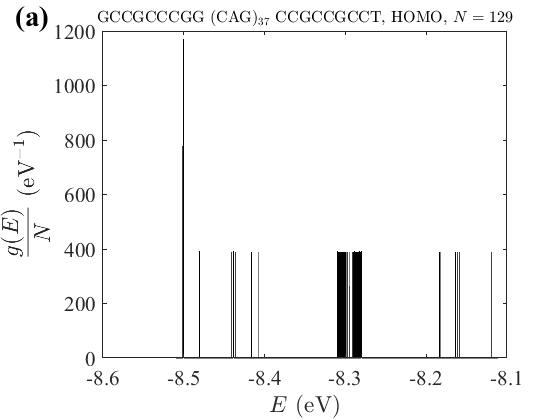}
\includegraphics[width=0.24\textwidth]{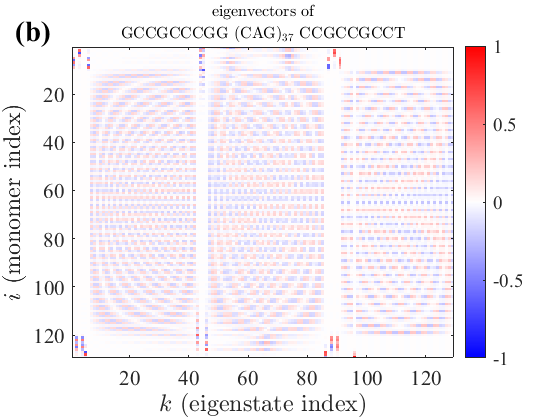}\\
\hspace{-0.5cm}
\includegraphics[width=0.24\textwidth]{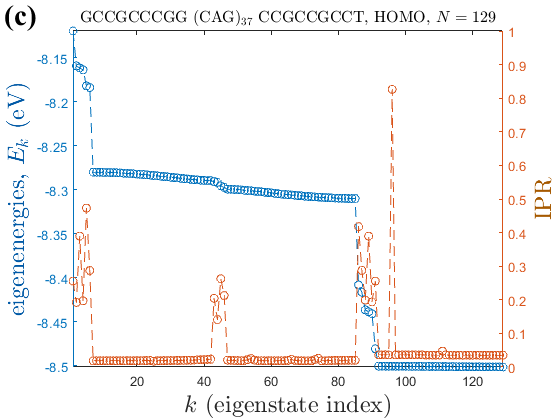}
\includegraphics[width=0.24\textwidth]{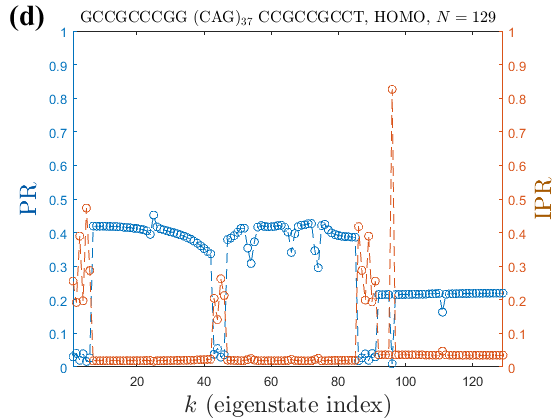}
\caption{Spinocerebellar Ataxia 7, for  $n=37$ ($N=129$): (a) density of States (DOS), (b) eigenvectors, (c) eigenenergies $E_k$ with inverse participation ratio (IPR), and (d) participation ratio (PR) with IPR.}
\label{fig:Ataxia7AllTIn37}
\end{figure}

For Spinocerebellar Ataxia 7, the time-independent features are illustrated Fig.~\ref{fig:Ataxia7AllTIn37} ($n=37$, $N=129$) and in Fig.~\ref{fig:Ataxia7AllTIn3} ($n=3$, $N=27$). Again, the periodicity of the \texttt{CAG} repeats promotes extended states. The primers introduce disorder, limiting the extension of states toward the edges, with a slightly stronger effect at the reverse primer. Increasing the number of repeats $n$, enhances the periodicity. Similarly, increasing the number of \texttt{CAG} repeats produces a more structured DOS profile.

We have studied systematically the electronic structure of holes, using the HOMOs of base pairs within the coarse-grained TB wire model.
The periodic CAG region promotes extended states, while the primers disrupt periodicity, preventing expansion of the states beyond the periodic segment. This suggests that the periodic CAG region facilitates hole transfer, whereas primers act as barriers that confine them.
Qualitatively similar electronic structure
characteristics are expected for extra electrons (using the LUMOs of base pairs), and consequently to a large extent for excitons or other quasi-particles. Increasing the number of repeats, $n$, strengthens periodicity, which in turn yields smoother spectra, higher PR values, and a more structured DOS. This trend is robust across all four CAG-repeat diseases, indicating that the length of the repeat region is a key factor in modulating electronic properties.

\subsection{Time-dependent}
\label{subsec:TD}
\begin{figure}[!h]
\centering
\hspace{-0.5cm}
\includegraphics[width=0.24\textwidth]{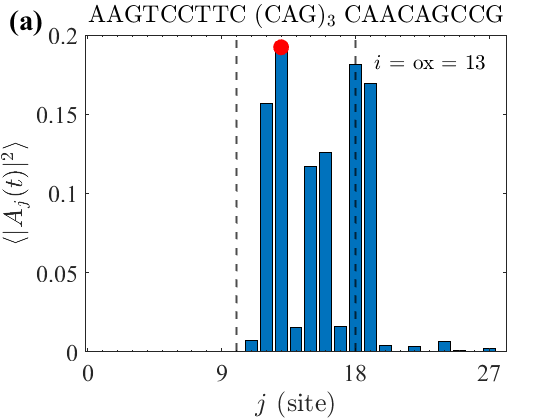}
\includegraphics[width=0.24\textwidth]{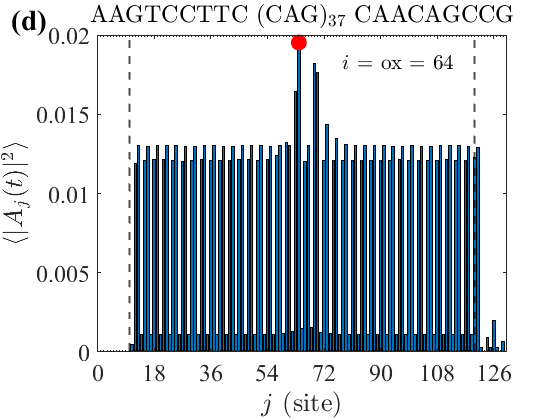}\\
\hspace{-0.5cm}
\includegraphics[width=0.24\textwidth]{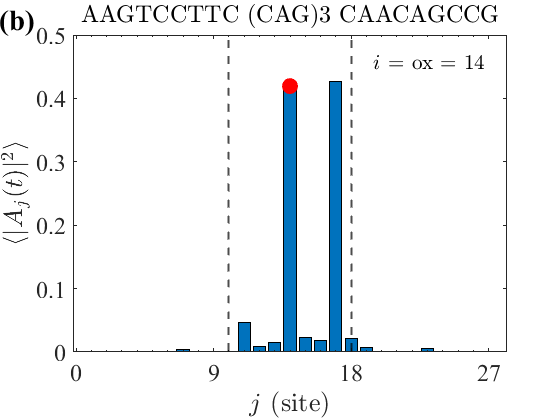}
\includegraphics[width=0.24\textwidth]{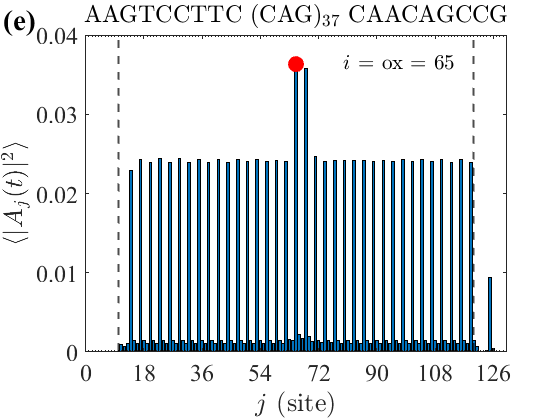}\\
\hspace{-0.5cm}
\includegraphics[width=0.24\textwidth]{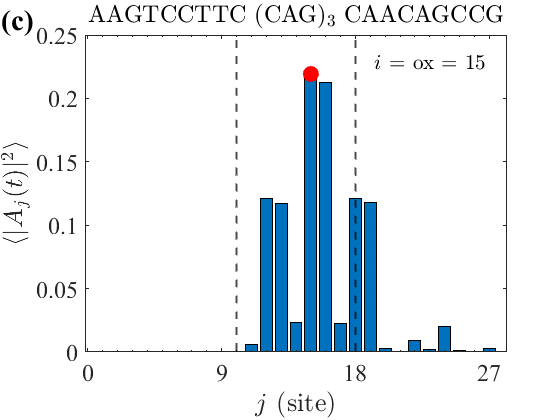}
\includegraphics[width=0.24\textwidth]{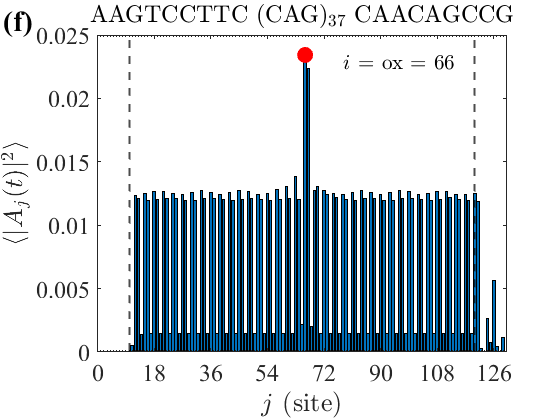}
\caption{Huntington’s disease. Mean over time probabilities for: 
$n=3$ ($N=27$) [left column], and 
$n=37$ ($N=129$) [right column]. 
The site of initial oxidation is denoted by the red filled circle. The vertical dashed lines indicate the boundaries between the central periodic region and the primers.
First row: Initial oxidation at the \texttt{C-G} base pair of the central trimer. 
Second row: Initial oxidation at the \texttt{A-T} base pair of the central trimer.
Third row:  Initial oxidation at the \texttt{G-C} base pair of the central trimer.}
\label{fig:HuntingtonTDmeanprob}
\end{figure}

\begin{figure}[!h]
\centering
\hspace{-0.5cm}
\includegraphics[width=0.24\textwidth]{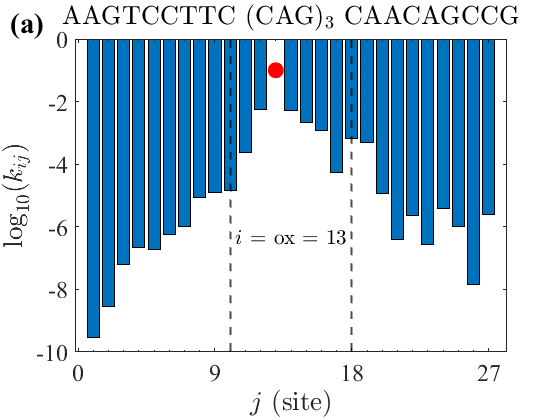}
\includegraphics[width=0.24\textwidth]{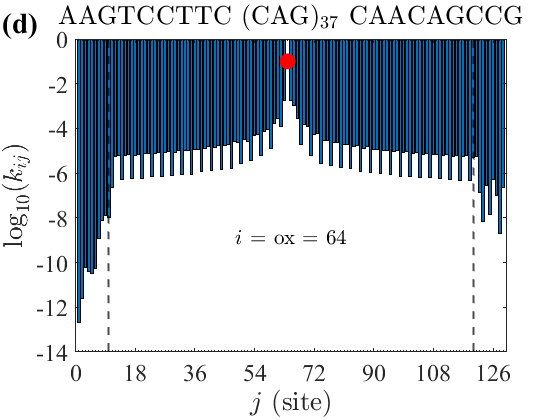}\\
\hspace{-0.5cm}
\includegraphics[width=0.24\textwidth]{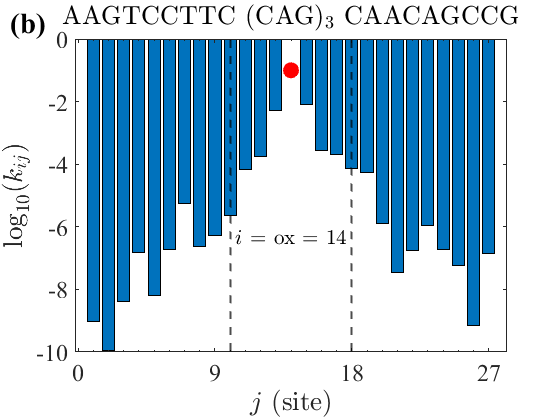}
\includegraphics[width=0.24\textwidth]{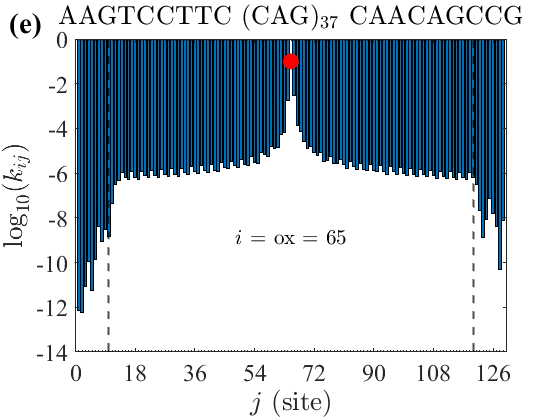}\\
\hspace{-0.5cm}
\includegraphics[width=0.24\textwidth]{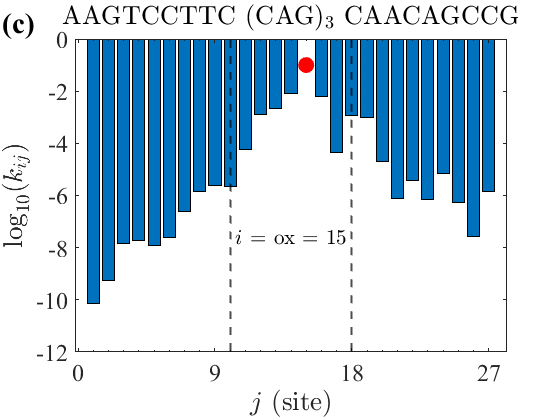}
\includegraphics[width=0.24\textwidth]{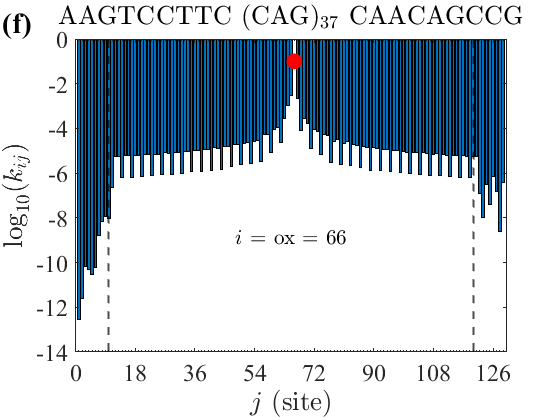}\\
\caption{Huntington’s disease. Mean transfer rates for: $n=3$ ($N=27$) [left column], and 
$n=37$ ($N=129$) [right column]. 
The site of initial oxidation is denoted by the red filled circle. The vertical dashed lines indicate the boundaries between the central periodic region and the primers.
First row: Initial oxidation at the \texttt{C-G} base pair of the central trimer. 
Second row: Initial oxidation at the \texttt{A-T} base pair of the central trimer.
Third row:  Initial oxidation at the \texttt{G-C} base pair of the central trimer.}
\label{fig:HuntingtonTDkij}
\end{figure}

Let us now examine what happens after an initial oxidation at the central \texttt{CAG} trimer. There are many other cases of oxidation that one might think of, but we think this case is sufficient to understand qualitatively what happens.
Fig.~\ref{fig:HuntingtonTDmeanprob} and Fig.~\ref{fig:HuntingtonTDkij} are devoted to the Huntington's disease. We introduce an initial oxidation at each of the three base pairs of the central \texttt{CAG} trimer, namely at 
\texttt{C-G} [panels (a) and (d)], \texttt{A-T} [panels (b) and (e)], and \texttt{G-C} [panels (c) and (f)]. 

In Fig.~\ref{fig:HuntingtonTDmeanprob} we present the mean over time probabilities of finding the hole at each monomer, and in Fig.~\ref{fig:HuntingtonTDkij} we present the mean transfer rates from the initial oxidation site to all other sites. These plots show that for all cases of initial oxidation within the central trimer, the hole is primarily distributed within the periodic segment. In contrast, in the non-periodic primers, both mean over time probabilities and mean transfer rates are negligible. As the sequence length increases, both the mean over time probability of the hole at each monomer of the periodic region and the mean transfer rates decrease, as expected, since the hole is distributed over a larger number of monomers. The transfer dynamics depend on the base pair at which the initial oxidation occurs. For oxidation at \texttt{C-G} or \texttt{G-C} pairs, the hole primarily moves between neighboring \texttt{C-G} and \texttt{G-C} pairs, while transitions to \texttt{A-T} pairs are suppressed. In contrast, for oxidation at \texttt{A-T} pairs, the hole primarily propagates between neighboring \texttt{A-T} pairs, with limited transfer to \texttt{C-G} or \texttt{G-C} pairs. The periodic \texttt{CAG} region supports delocalized states, while the primers act as barriers that suppress charge transfer.

The time-dependent dynamics for the remaining three diseases are shown in Figs.
\ref{fig:KennedyTDmeanprob}-\ref{fig:KennedyTDkij} (for Kennedy's disease),
\ref{fig:Ataxia6TDmeanprob}-\ref{fig:Ataxia6TDkij} (for Spinocerebellar Ataxia 6) 
and \ref{fig:Ataxia7TDmeanprob}-\ref{fig:Ataxia7TDkij} (for Spinocerebellar Ataxia 7).
As for Huntington's disease, we initialize the hole at each of the three base-pair types within the central \texttt{CAG} trimer: \texttt{C-G}, \texttt{A-T}, and \texttt{G-C} (rows 1–3, respectively). For each disease, we present both the mean over time probabilities and the mean transfer rates, for small and large number of repeats, $n$.

The results are qualitatively consistent across all diseases: the hole remains largely confined to the central periodic region and, for Kennedy's disease, also to the forward primer, which closely resembles the periodic segment. The reverse primer consistently acts as a strong barrier, with negligible probability or transfer rates beyond it. As the number of repeats increases, the probability per monomer decreases due to the larger available space, while the overall extent of delocalization grows. The base-pair selectivity observed for Huntington's disease is also recovered here: oxidation at \texttt{C-G} or \texttt{G-C} sites preferentially populates \texttt{C-G}/\texttt{G-C} sites, whereas oxidation at \texttt{A-T} sites primarily transfers among \texttt{A-T} (and \texttt{T-A}, where applicable) sites. These findings reinforce the general picture that the periodic \texttt{CAG} region supports extended states and facilitates charge transfer, while the primers disrupt periodicity, preventing expansion beyond the periodic segment.

Compared to our previous studies, the present work provides a systematic comparison across four diseases, a detailed analysis of the role of primers, and a complete time-dependent characterization of hole transfer, thereby establishing general trends for \texttt{CAG}-repeat sequences. In this respect, it complements Ref.~\cite{MLS:2023}, which explored charge transport for the same set of diseases.

\subsection{Connection to pathogenicity}
\label{subsec:pathogenicity}
We have discussed short, non-pathogenic reference sequences ($n=3$) and sequences with pathological (e.g., $n=37$ for Huntington's disease) \texttt{CAG} repeat numbers. Is there a quantitative relationship between $n$, electronic structure, oxidation dynamics, and disease emergence? To investigate this issue, we systematically calculated the mean over time probabilities of finding the hole, after an initial oxidation at the central trimer, as a function of $n$ and examined whether any clear transition occurs near clinically relevant repeat lengths. In other words, we have searched for a demonstration of whether the studied observables distinguish pathogenic from non-pathogenic repeat lengths. 

\begin{figure}[h]
\centering
\includegraphics[width=0.45\textwidth]{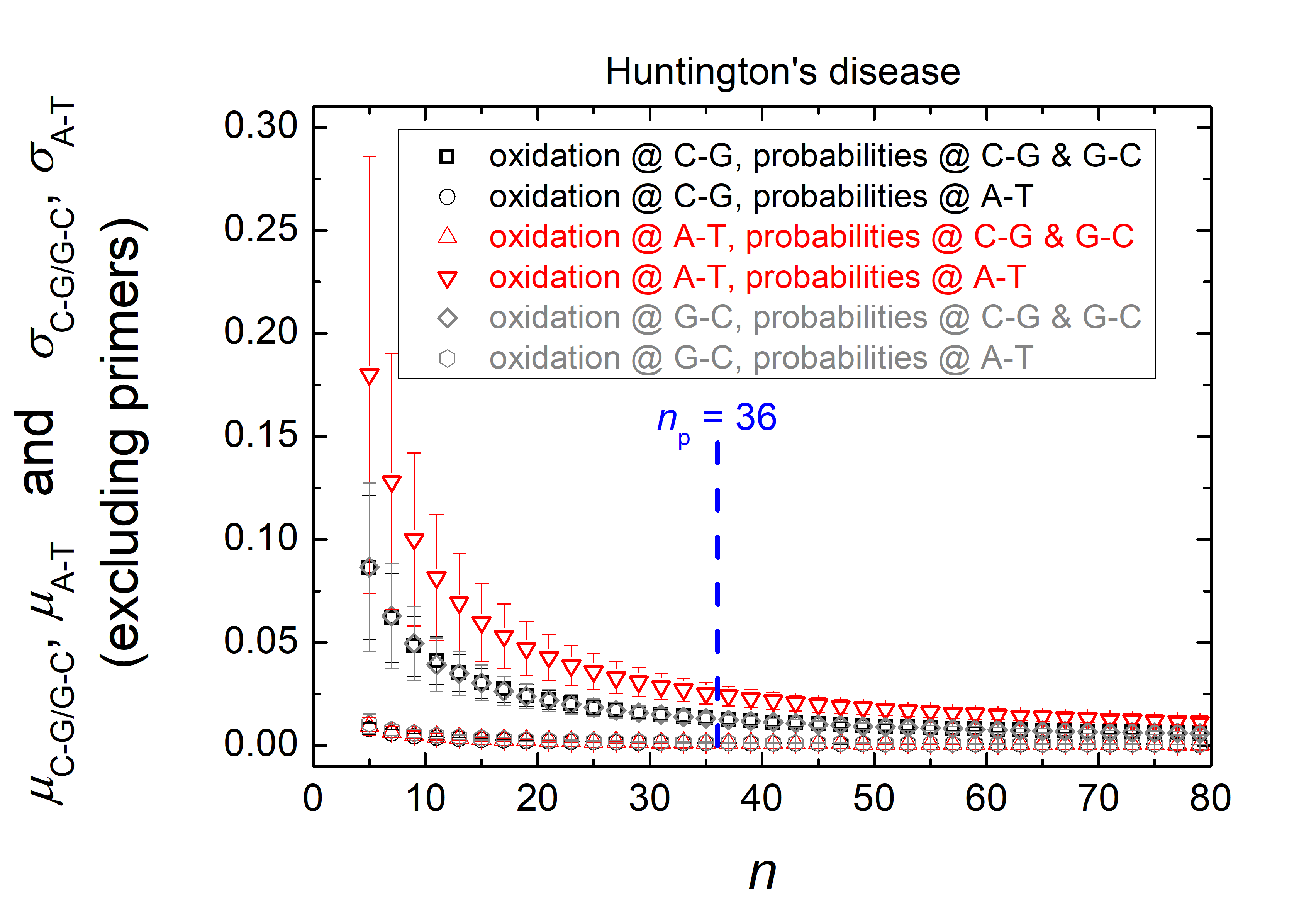}
\caption{Huntington's disease. After excluding primers, which act as barriers: Spatial average of mean over time probabilities at \texttt{C-G} or \texttt{G-C} sites, $\mu_\texttt{C-G/G-C} := \overline{\langle |C_j(t)|^2\rangle_\texttt{C-G or G-C}}$, and at \texttt{A-T} sites $\mu_\texttt{A-T} := \overline{\langle |C_\mu(t)|^2\rangle_\texttt{A-T}}$ as well as their corresponding standard deviations, $\sigma_\texttt{C-G/G-C}$ and $\sigma_\texttt{A-T}$, respectively, as  functions of the number of repeats, $n$. Initial oxidation is performed at the \texttt{C-G} or \texttt{A-T} or \texttt{G-C} base pair of the central (CAG) trimer.}
\label{fig:HuntingtonMeanOfMeans}
\end{figure}

Since hole delocalization may impede recognition by repair proteins like OGG1, which scans the sequence to detect and repair oxidative damage, particularly to guanine (e.g., 8-oxoguanine), let's see when scanning becomes ineffective.
We calculated the mean over time probabilities $\langle |C_j(t)|^2\rangle$ at each site $j$, then excluded the primers and calculated the spatial average of $\langle |C_j(t)|^2\rangle$ exclusively for \texttt{C-G} or \texttt{G-C} sites as well as for \texttt{A-T} sites [as we read them in the $5'-3'$ direction]. In other words, we evaluated the spatial averages $\mu_\texttt{C-G/G-C} := \overline{\langle |C_j(t)|^2\rangle_\texttt{C-G or G-C}}$ and $\mu_\texttt{A-T} := \overline{\langle |C_j(t)|^2\rangle_\texttt{A-T}}$, along with  standard deviations, $\sigma_\texttt{C-G/G-C}$ and $\sigma_\texttt{A-T}$, respectively.
Oxidation is always performed at the central trimer, at \texttt{C-G} or \texttt{A-T} or \texttt{G-C}, respectively. These spatial averages and standard deviations are shown in Fig.~\ref{fig:HuntingtonMeanOfMeans}.

We observe that these spatial averages decrease smoothly with $n$, behaving similarly across the four genes as the hole spreads over a longer tract. As the sequence length approaches the respective clinical threshold range ($n_p$)  these probabilities dilute: 
(i) for oxidation at \texttt{C-G} or \texttt{G-C} sites, $\mu_\texttt{C-G/G-C} \approx 0.01$ and $\mu_\texttt{A-T} \approx 0.00$ and (ii) for oxidation at \texttt{A-T} sites $\mu_\texttt{A-T} \approx 0.02$ and $\mu_\texttt{C-G/G-C}\approx 0.00$. 
Similar is the situation for the other three diseases, cf. Appendix~\ref{sec:TDrest}, Fig.~\ref{fig:KennedyMeanOfMeans}, Fig.~\ref{fig:Ataxia6MeanOfMeans}, Fig.~\ref{fig:Ataxia7MeanOfMeans}.
Although the range of pathological repetitions is different (cf. Table~\ref{tab:gene_data}), we depict, for homogeneity and clarity reasons, up to $n = 80$, for all diseases.  
Hence, the pathological threshold seems to be connected to inability of recognizing the oxidation damage due to expansion of the hole to many base pairs. At this point, the system essentially behaves as an extended, periodic one-dimensional chain, rendering the confining influence of the primers negligible. Rather than displaying a sharp physical feature at a specific threshold, the $n$-dependence exhibits a smooth crossover that correlates with repeat expansion. We hypothesize that this extensive, continuous delocalization of the hole effectively "masks" the initial oxidative damage, preventing scanning repair proteins, such as OGG1, from successfully pinpointing and repairing localized 8-oxoguanine lesions.
Although our current model evaluates the coherent hole transfer in structurally canonical sequences rather than explicitly incorporating the altered on-site energies of a permanent chemical lesion (like 8-oxoG) and possible modifications of relevant interaction integrals, this continuous dilution of the charge probability could severely disrupt the scanning mechanisms of repair proteins such as OGG1. It is also worth noting that the biological role of OGG1 is complex and debated, e.g., it has been reported that 
OGG1 initiates age-dependent CAG trinucleotide expansion in somatic cells~\cite{Kovtun:2007}.

\begin{figure}[h]
\centering
\includegraphics[width=0.45\textwidth]{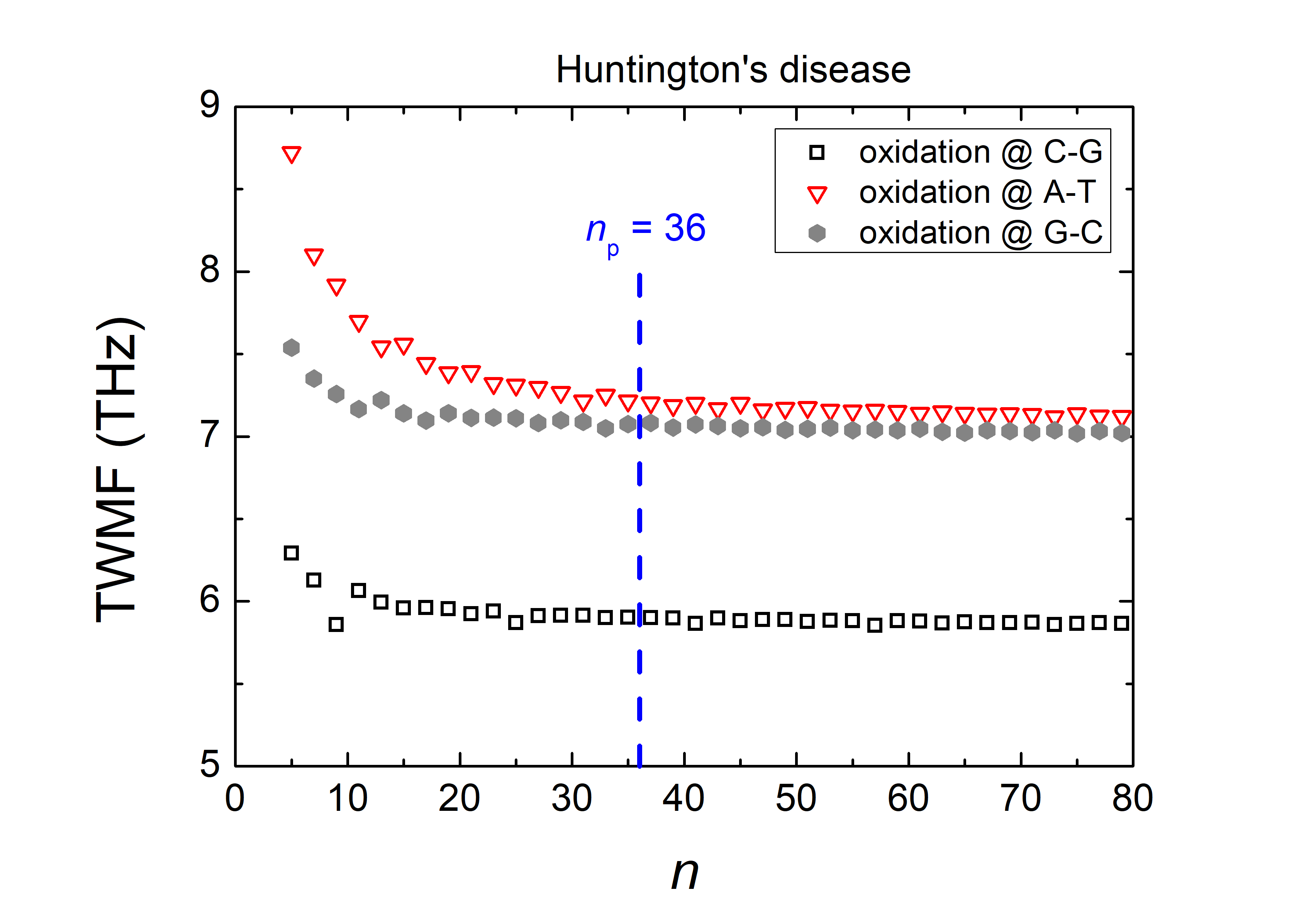}
\caption{Huntington's disease. Total Weighted Mean Frequency (TWMF) as a function of the number of repeats, $n$. Initial oxidation is performed at the \texttt{C-G} or \texttt{A-T} or \texttt{G-C} base pair of the central \texttt{CAG} trimer.}
\label{fig:HuntingtonTWMF}
\end{figure}

\begin{figure}[h]
\centering
\includegraphics[width=0.45\textwidth]{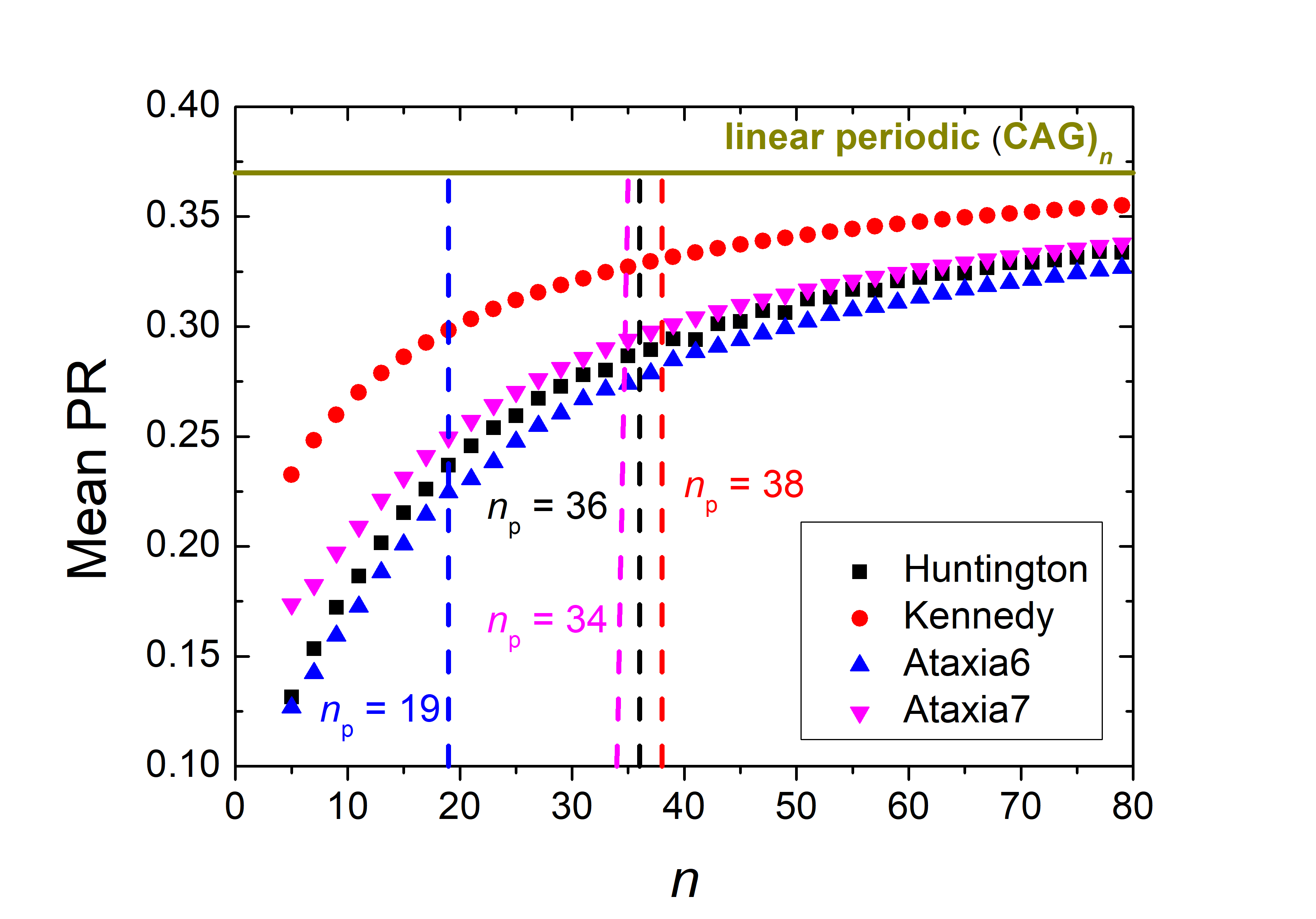}
\caption{Mean Participation Ratio (PR) as a function of the number of repeats, $n$, for the four diseases.}
\label{fig:MeanPR}
\end{figure}

To further correlate physical observables with pathogenic thresholds, we calculated, without excluding the primers, the Total Weighted Mean Frequency (TWMF) and the Mean Participation Ratio (PR) as a function of the \texttt{CAG} repeat number, $n$. While short sequences are strongly confined by the non-periodic flanking primers, increasing $n$ enables charge delocalization, leading to a steep increase in Mean PR and a sharp drop in TWMF. Both observables exhibit a smooth crossover as a function of $n$, entering their long-chain asymptotic regime in the vicinity of the respective clinical thresholds $n_p$. The TWMF for the Huntington's disease is shown in Fig.~\ref{fig:HuntingtonTWMF}. For the other diseases cf. Figs.~\ref{fig:KennedyTWMF}, ~\ref{fig:Ataxia6TWMF},~\ref{fig:Ataxia7TWMF}, in the Appendix.
The Mean of the Participation Ratio (PR) as a function of the number of repeats, $n$, for the four diseases, is presented in Fig. \ref{fig:MeanPR}.

Here we mention that an infinite linear (non cyclic) periodic sequence (\texttt{CAG)}$_n$ has a Mean PR of approximately $ \left( \frac{2N}{3} \frac{4}{9} + \frac{N}{3} \frac{2}{9} \right) / N = \frac{10}{27} \approx 0.37$.  Hence, Fig.~\ref{fig:MeanPR} shows that the four sequences approach the Mean PR of the corresponding infinite linear (non-cyclic) periodic \texttt{CAG} sequence.

Figures~\ref{fig:HuntingtonMeanOfMeans}-\ref{fig:HuntingtonTWMF} (as well as Figs. ~\ref{fig:KennedyMeanOfMeans}-\ref{fig:KennedyTWMF},  
~\ref{fig:Ataxia6MeanOfMeans}-\ref{fig:Ataxia6TWMF},  
~\ref{fig:Ataxia7MeanOfMeans}-\ref{fig:Ataxia7TWMF}) and Fig.~\ref{fig:MeanPR} illustrate three related but not redundant measures: local (spatial probability), dynamical (frequency content), static (localization/delocalization) for the connection to pathogenicity.


\subsection{An incoherent approach is not enough}
\label{subsec:incoherent}
For an incoherent hopping process, the charge transfer rate between sites $i$ and $j$ can be estimated using the semiclassical Marcus theory \cite{MarcusSutin:1985}, i.e.,
\begin{equation}
k_{ij} = \frac{2\pi}{\hbar} |t_{ij}|^2 
\frac{1}{\sqrt{4\pi \lambda k_B T}}
\exp\left[-\frac{(\Delta E + \lambda)^2}{4\lambda k_B T}\right],
\label{eq:Marcus}
\end{equation}
where $k_{ij}$ is the hopping rate from site $i$ to site $j$, $t_{ij}$ is the coupling (or interaction or ``hopping'' or ``transfer'') integral between sites $i$ and $j$, $\lambda$ is the total reorganization energy, comprising inner-sphere (vibrational) and outer-sphere (solvent) contributions,
$T$ is the absolute temperature, and $\Delta E = E_j - E_i$ is the driving force, i.e., the energy difference between the final and initial sites. Typical reorganization energies for hole transfer in DNA are of the order of $\lambda\approx0.5$--$1.0$ eV, depending on the base pairs involved and the  environment. Values of $\approx$ $0.7$--$0.9$ eV have been reported for G-C base pairs, whereas A-T base pairs typically exhibit somewhat smaller reorganization energies of about $0.4$--$0.6$ eV \cite{Voityuk:2006,Khan:2013}. 

\begin{figure}[!h]
	\centering
	\hspace{-0.5cm}
	\includegraphics[width=0.44\textwidth]{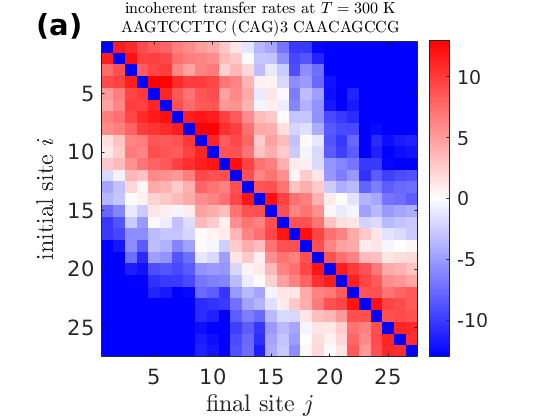}
	\includegraphics[width=0.44\textwidth]{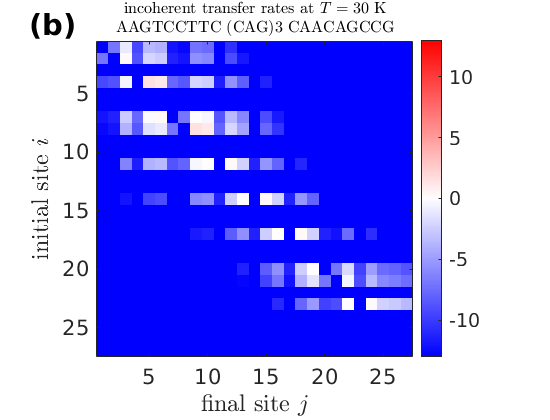}\\
	\caption{Huntington's disease. Incoherent transfer rates, shown as $\log_{10}k_{ij}$, using Marcus theory for $n=3$ ($N=27$): (a) $T = 300$ K, (b) $T = 30$ K. At lower $T$, the incoherent rates become negligible (note the colorbar scale).
	Rates are computed for all pairs $(i, j)$; near-neighbor contributions dominate because the TB couplings decay rapidly with distance; longer-range contributions, included via the super-exchange coupling, are smaller. 
	The primer regions (indicated by the vertical dashed lines in the corresponding coherent plots) show no signature of primer-induced suppression, indicating that the barrier is not of thermal origin.}
	\label{fig:HuntingtonTDkijMarcus}
\end{figure}

Fig.~\ref{fig:HuntingtonTDkijMarcus} tests whether an incoherent theory can capture the existence of the periodic barrier. Since there is no universal agreement (neither on inner reorganization energies nor on solvent contribution), we have used as representative values $\lambda_\texttt{G-C} = 0.8$ eV for \texttt{G-C} or \texttt{C-G} and $\lambda_\texttt{A-T} = 0.5$ eV for \texttt{A-T} or \texttt{T-A}, assigning to each hopping process the arithmetic mean of the reorganization energies of the two participating sites, a common approximation for heterogeneous hopping between localized states. Changing these values within acceptable limits does not change the general picture.

This Marcus-type hopping model, incorporating base-pair-dependent reorganization energies and electronic couplings derived from the same tight-binding Hamiltonian, 
does not reproduce the primer-induced suppression of charge transfer observed in the coherent calculations. Fig.~\ref{fig:HuntingtonTDkijMarcus}
shows only two temperatures, but this 
behavior persists across all temperatures. Lowering the temperature simply reduces all hopping rates, eventually rendering them negligibly small, without generating the primer-induced suppression observed in the coherent calculations. 
This indicates that the barrier is of coherent quantum-mechanical origin, arising from the periodic structure of the \texttt{CAG} repeats, and cannot be explained by thermally activated hopping.


\section{Conclusion}
\label{sec:Conclusion}
The four genes exhibit qualitatively similar behavior with respect to the studied physical properties. The periodicity of the \texttt{CAG} repeats in the central part of the sequence promotes hole delocalization. This means that the hole spreads relatively homogeneously along the \texttt{CAG} region, as indicated by the higher participation ratios. As the number of repeats increases, the periodicity is enhanced, and the energy subbands become more clearly distinguishable.

The primers at the ends of the sequence introduce disorder, which hinders further delocalization of the hole. Consequently, the probability of finding the hole in these regions is low. This barrier is of purely coherent quantum-mechanical origin, arising from the periodic structure of the \texttt{CAG} repeats and the disruption of periodicity at the primers, and cannot be explained by thermally activated hopping.
This disruption of periodicity generates fluctuations in the energy subbands and introduces additional energy levels. In Kennedy’s disease, the forward primer has a sequence similar to the repeated segment (with inverted \texttt{A-T} vs. \texttt{T-A} base pairs), allowing the hole to spread into that region. In contrast, the reverse primer, which differs significantly from the repeated segment, inhibits hole propagation in that region.

The mean transfer rates and the probability of finding the hole at each base pair are highest within the periodic \texttt{CAG} segment and very low in the primer regions. As the sequence length increases, both metrics decrease, since the hole spreads over a greater distance.

As \texttt{G-C} (or \texttt{C-G}) has the lowest ionization potential, oxidation is more probable at \texttt{G-C} (or \texttt{C-G}) sites. However, within our approach, we can introduce the initial oxidation, at any site. A hole generated at a \texttt{C-G} or \texttt{G-C} base pair is primarily transferred through other \texttt{C-G} and \texttt{G-C} pairs, while transfer through \texttt{A-T} pairs is much less probable. Similarly, a hole generated at an \texttt{A-T} base pair is mainly transferred through \texttt{A-T} pairs, with limited transfer through \texttt{C-G} and \texttt{G-C} pairs. This behavior reflects the hole's preference to move coherently through base pairs with similar on-site energies.

The physical behavior of holes in \texttt{CAG} repeat segments influences gene biology and can contribute to pathological conditions. As mentioned in Section~\ref{sec:Introduction}, mutated polyQ sequences lead to cellular stress, which produces ROS capable of oxidizing DNA bases such as guanine, forming 8-oxoguanine (8-oxoG). Hole delocalization may impede recognition by repair proteins like OGG1, which scans the sequence to detect and repair oxidative damage, particularly to guanine (e.g., 8-oxoguanine).

With increasing the \texttt{CAG} repeat number, the hole becomes progressively more delocalized over the periodic region, as reflected by the decrease of the TWMF and the increase of the Mean Participation Ratio. 
The TWMF exhibits a smooth crossover, transitioning toward its long-chain asymptotic behavior as the periodic tract expands through the clinically pathogenic repeat range. Mean PR progressively approaches the asymptotic value of $10/27 \approx 0.37$, characteristic of an infinite linear periodic \texttt{CAG}$_n$ sequence. 
These results indicate a close correlation between repeat expansion and the transition toward an extended, periodic electronic regime. An incoherent Marcus-type hopping description, in contrast, does not reproduce the primer-induced suppression of charge transfer observed in the coherent calculations, supporting the interpretation that this barrier originates from the coherent quantum-mechanical structure of the finite sequence.

\bibliography{CAGnAfterAcceptance_arXiv}

\clearpage


%

\FloatBarrier

\appendix
\counterwithin{figure}{section}
\counterwithin{table}{section}
\renewcommand{\thefigure}{A\arabic{figure}}
\renewcommand{\thetable}{A\arabic{table}}

\section{Plots for sequences with 3 \texttt{CAG} repeats}
\label{sec:n3plots}
\begin{figure}[h]
\centering
\hspace{-0.5cm}
\includegraphics[width=0.24\textwidth]{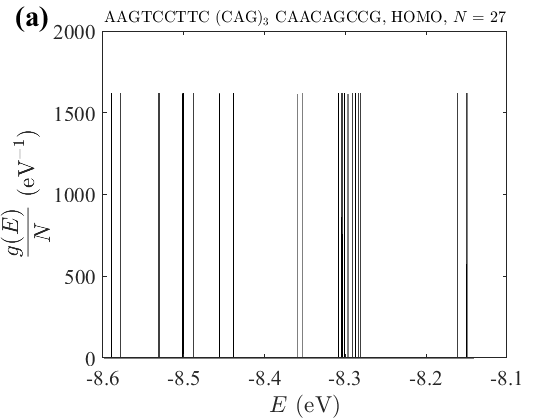}
\includegraphics[width=0.24\textwidth]{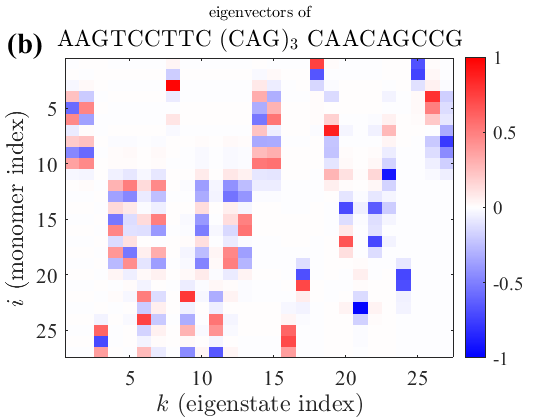}\\
\hspace{-0.5cm}
\includegraphics[width=0.24\textwidth]{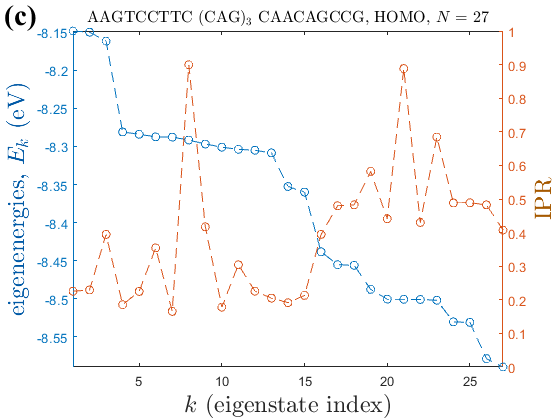}
\includegraphics[width=0.24\textwidth]{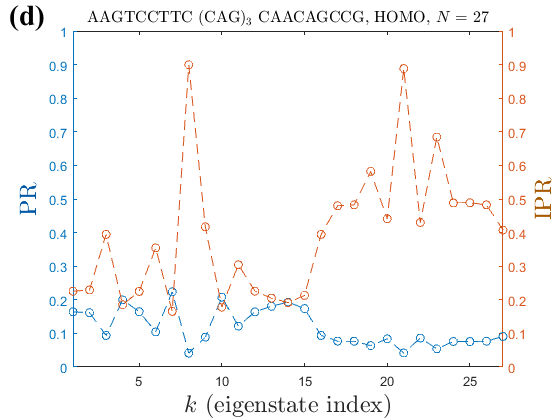}
\caption{Huntington’s disease, for $n=3$ ($N=27$): 
(a) density of States (DOS), (b) eigenvectors, (c) eigenenergies $E_k$ with inverse participation ratio (IPR), and (d) participation ratio (PR) with IPR.}
\label{fig:HuntingtonAllTIn3}
\end{figure}

\begin{figure}[h]
\centering
\hspace{-0.5cm}
\includegraphics[width=0.24\textwidth]{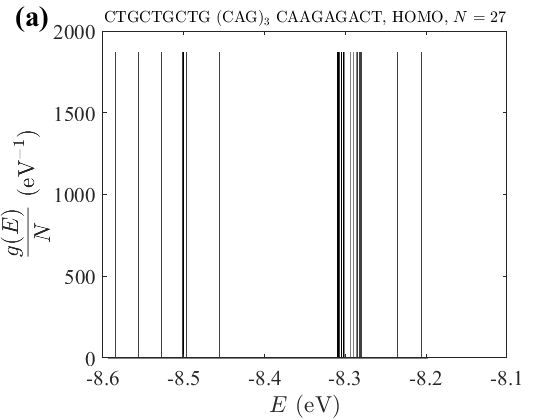}
\includegraphics[width=0.24\textwidth]{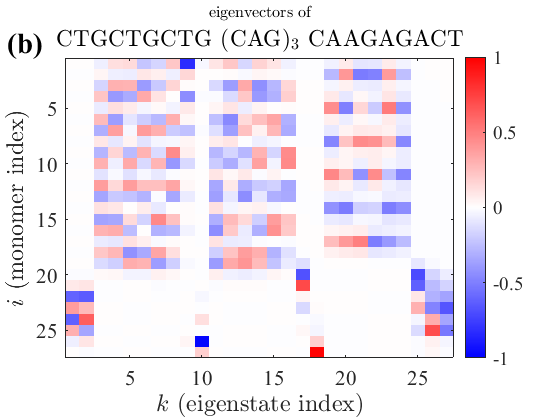}\\
\hspace{-0.5cm}
\includegraphics[width=0.24\textwidth]{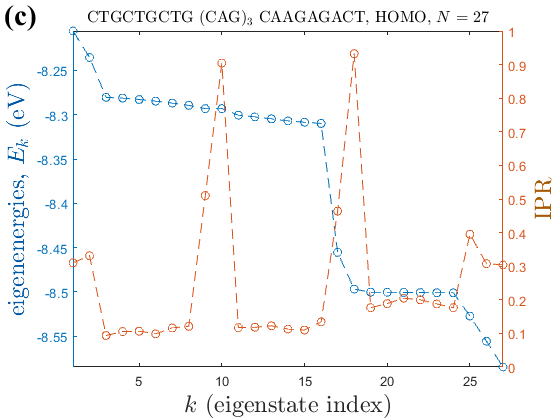}
\includegraphics[width=0.24\textwidth]{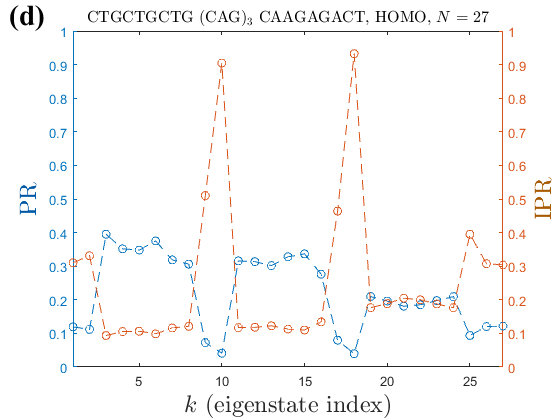}
\caption{Kennedy’s disease, for $n=3$ ($N=27$): 
(a) density of States (DOS), (b) eigenvectors, (c) eigenenergies $E_k$ with inverse participation ratio (IPR), and (d) participation ratio (PR) with IPR.}
\label{fig:KennedyAllTIn3}
\end{figure}

\begin{figure}[h]
\centering
\hspace{-0.5cm}
\includegraphics[width=0.24\textwidth]{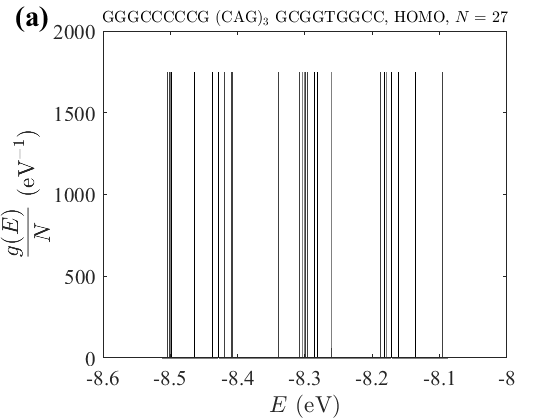}
\includegraphics[width=0.24\textwidth]{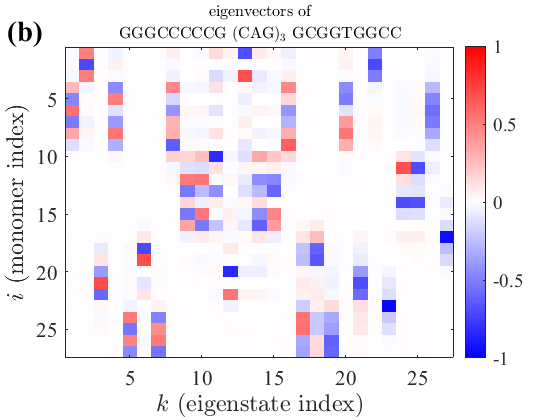}\\
\hspace{-0.5cm}
\includegraphics[width=0.24\textwidth]{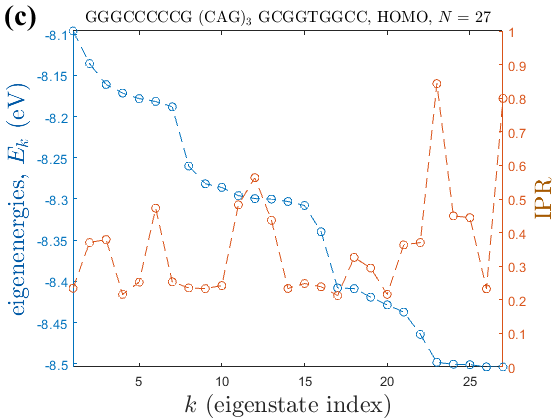}
\includegraphics[width=0.24\textwidth]{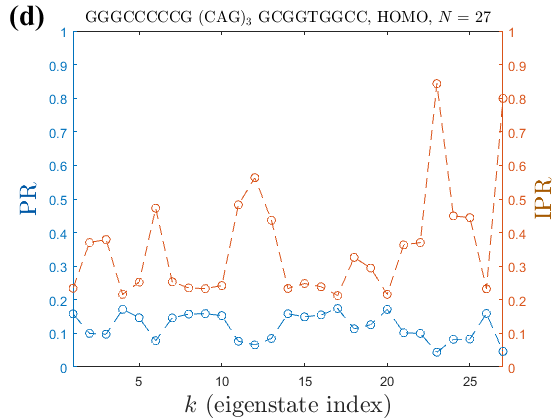}
\caption{Spinocerebellar Ataxia 6, for $n=3$ ($N=27$): 
(a) density of States (DOS), (b) eigenvectors, (c) eigenenergies $E_k$ with inverse participation ratio (IPR), and (d) participation ratio (PR) with IPR.}
\label{fig:Ataxia6AllTIn3}
\end{figure}

\begin{figure}[h]
\centering
\hspace{-0.5cm}
\includegraphics[width=0.24\textwidth]{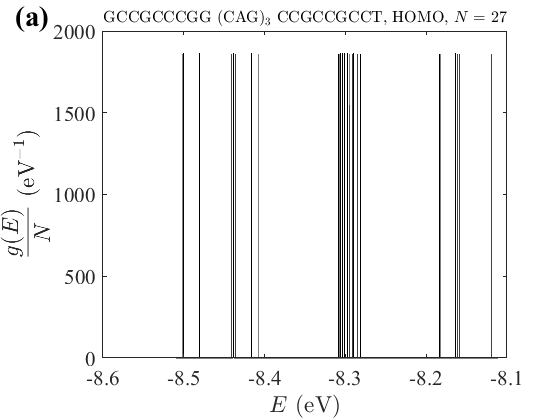}
\includegraphics[width=0.24\textwidth]{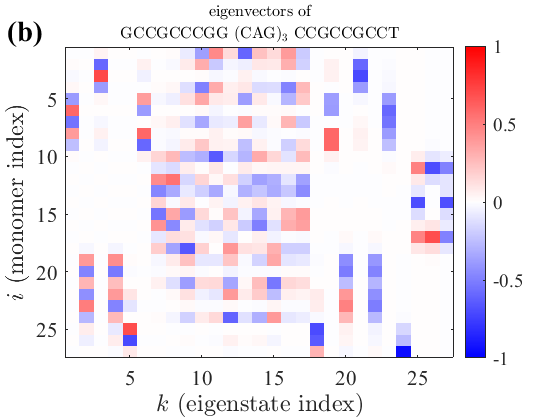}\\
\hspace{-0.5cm}
\includegraphics[width=0.24\textwidth]{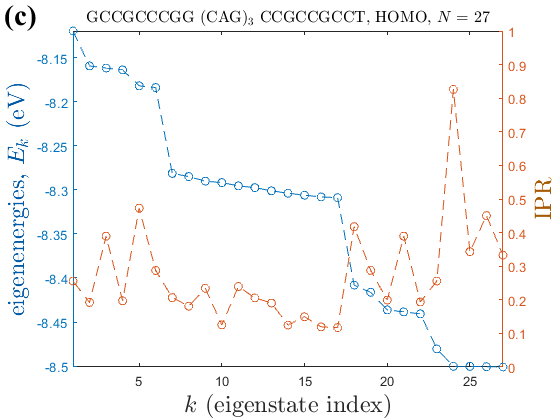}
\includegraphics[width=0.24\textwidth]{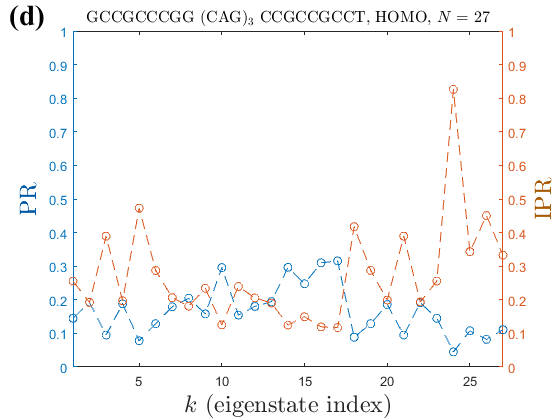}
\caption{Spinocerebellar Ataxia 7, for $n=3$ ($N=27$): 
(a) density of States (DOS), (b) eigenvectors, (c) eigenenergies $E_k$ with inverse participation ratio (IPR), and (d) participation ratio (PR) with IPR.}
\label{fig:Ataxia7AllTIn3}
\end{figure}

The following figures present various aspects of the electronic structure, at $n=3$, for Huntington's disease (Fig.~\ref{fig:HuntingtonAllTIn3}), Kennedy's disease (Fig.~\ref{fig:KennedyAllTIn3}), Spinocerebellar Ataxia 6 (Fig.~\ref{fig:Ataxia6AllTIn3}), and Spinocerebellar Ataxia 7 (Fig.~\ref{fig:Ataxia7AllTIn3}). At this small $n$, some features, particularly eigenvector components, can be readily seen.

\newpage

\section{Time-dependent results for Kennedy's disease and Spinocerebellar Ataxias 6 and 7}
\label{sec:TDrest}
\setcounter{figure}{0}
\renewcommand{\thefigure}{B\arabic{figure}}
Figs.~\ref{fig:KennedyTDmeanprob}, 
~\ref{fig:KennedyMeanOfMeans}, 
~\ref{fig:KennedyTDkij}, 
~\ref{fig:KennedyTWMF} 
are devoted to Kennedy's disease.
Figs.~\ref{fig:Ataxia6TDmeanprob},  ~\ref{fig:Ataxia6MeanOfMeans},  
~\ref{fig:Ataxia6TDkij},  
~\ref{fig:Ataxia6TWMF} 
are devoted to Spinocerebellar Ataxia 6. 
Figs.~\ref{fig:Ataxia7TDmeanprob},  ~\ref{fig:Ataxia7MeanOfMeans}, 
~\ref{fig:Ataxia7TDkij}, 
~\ref{fig:Ataxia7TWMF} 
are devoted to Spinocerebellar Ataxia 7.

\begin{figure}[h]
\centering
\hspace{-0.5cm}
\includegraphics[width=0.24\textwidth]{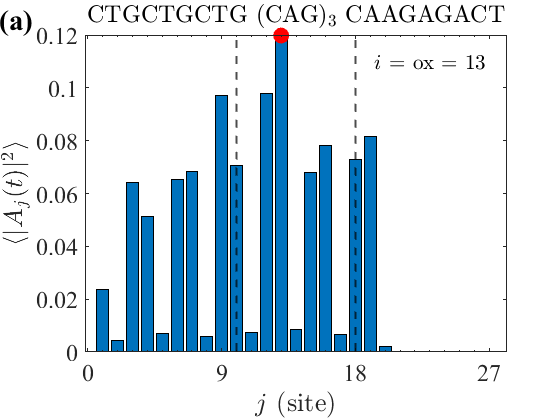}
\includegraphics[width=0.24\textwidth]{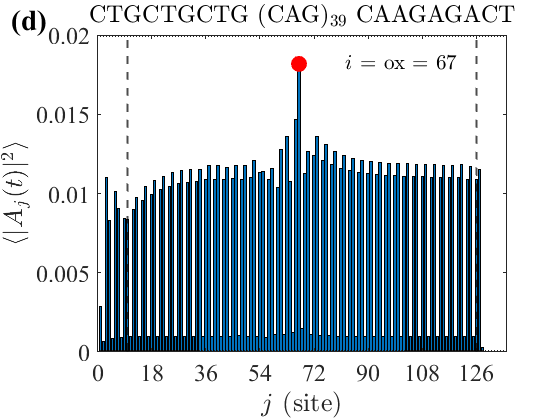}\\
\hspace{-0.5cm}
\includegraphics[width=0.24\textwidth]{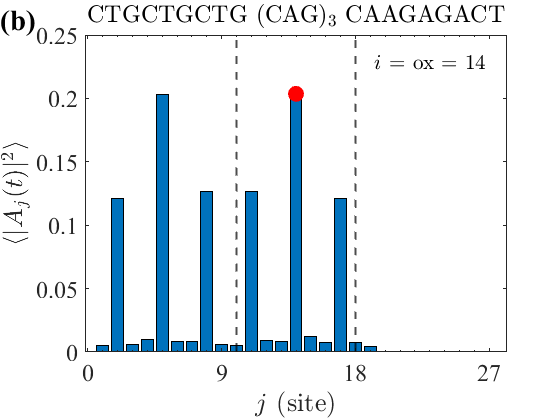}
\includegraphics[width=0.24\textwidth]{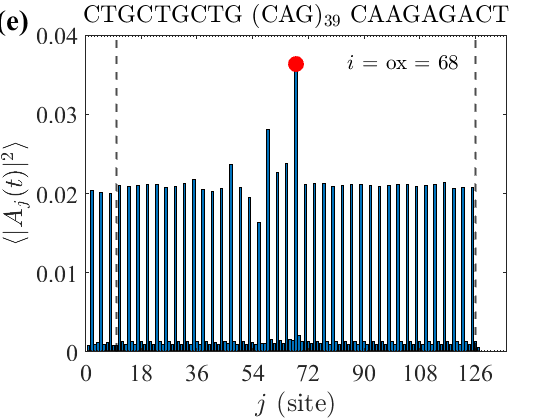}\\
\hspace{-0.5cm}
\includegraphics[width=0.24\textwidth]{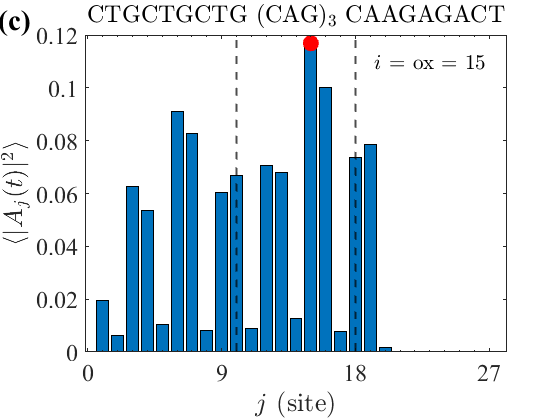}
\includegraphics[width=0.24\textwidth]{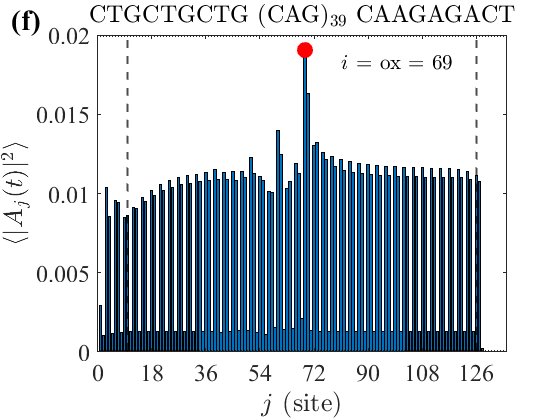}\\
\caption{Kennedy’s disease. Mean over time probabilities for: $n=3$ ($N=27$) [left column], and $n=39$ ($N=135$) [right column]. The site of initial oxidation is denoted by the red filled circle. The vertical dashed lines indicate the boundaries between the central periodic region and the primers.
First row: Initial oxidation at the \texttt{C-G} base pair of the central trimer. 
Second row: Initial oxidation at the \texttt{A-T} base pair of the central trimer.
Third row:  Initial oxidation at the \texttt{G-C} base pair of the central trimer.}
\label{fig:KennedyTDmeanprob}
\end{figure}

\begin{figure}[h]
\centering
\includegraphics[width=0.45\textwidth]{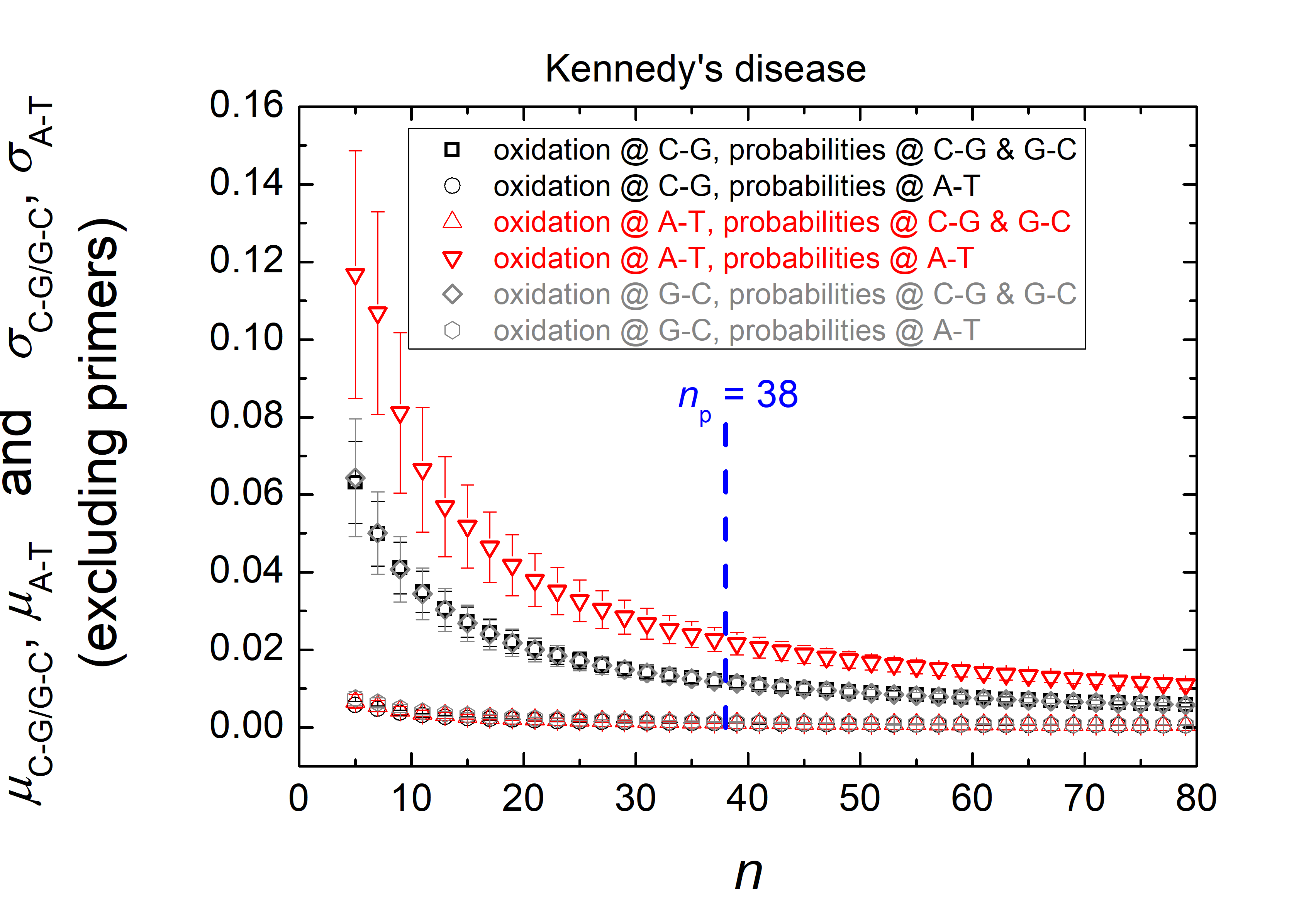}
\caption{Kennedy's disease. After excluding primers, which act as barriers: Spatial average of mean over time probabilities at \texttt{C-G} or \texttt{G-C} sites, $\mu_\texttt{C-G/G-C} := \overline{\langle |C_j(t)|^2\rangle_\texttt{C-G or G-C}}$, and at \texttt{A-T} sites $\mu_\texttt{A-T} := \overline{\langle |C_\mu(t)|^2\rangle_\texttt{A-T}}$ as well as their corresponding standard deviations, $\sigma_\texttt{C-G/G-C}$ and $\sigma_\texttt{A-T}$, respectively, as functions of the number of repeats, $n$. Initial oxidation is performed at the \texttt{C-G} or \texttt{A-T} or \texttt{G-C} base pair of the central \texttt{CAG} trimer.}
\label{fig:KennedyMeanOfMeans}
\end{figure}

\begin{figure}[h]
\centering
\hspace{-0.5cm}
\includegraphics[width=0.24\textwidth]{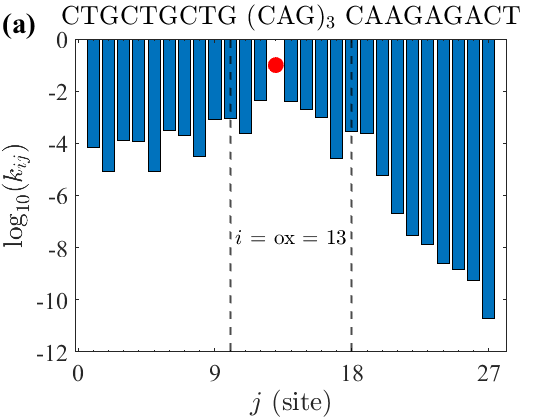}
\includegraphics[width=0.24\textwidth]{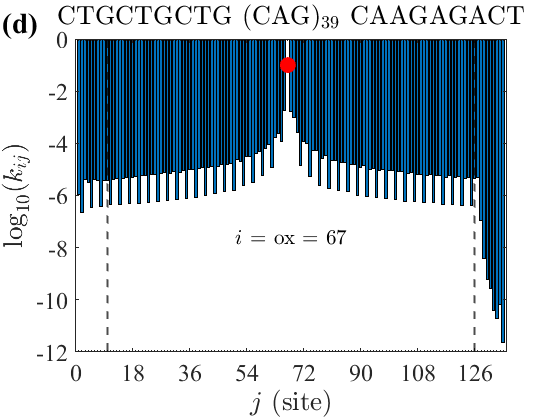}\\
\hspace{-0.5cm}
\includegraphics[width=0.24\textwidth]{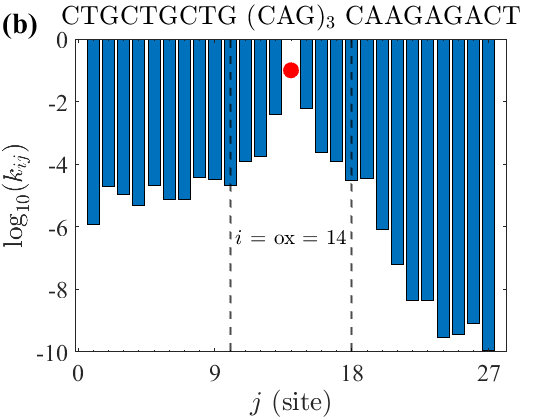}
\includegraphics[width=0.24\textwidth]{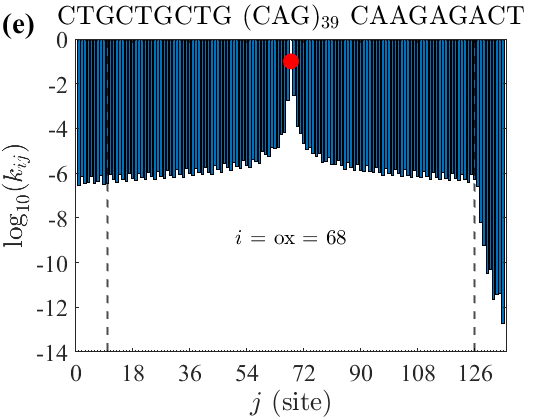}\\
\hspace{-0.5cm}
\includegraphics[width=0.24\textwidth]{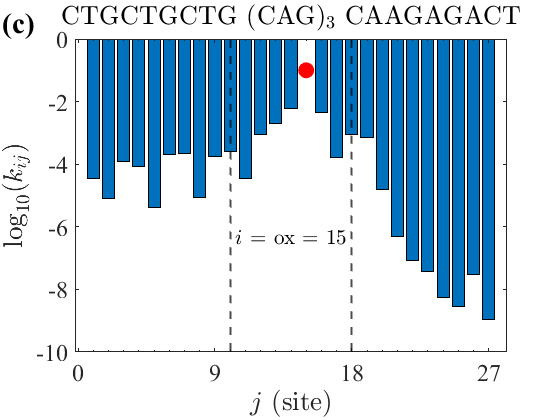}
\includegraphics[width=0.24\textwidth]{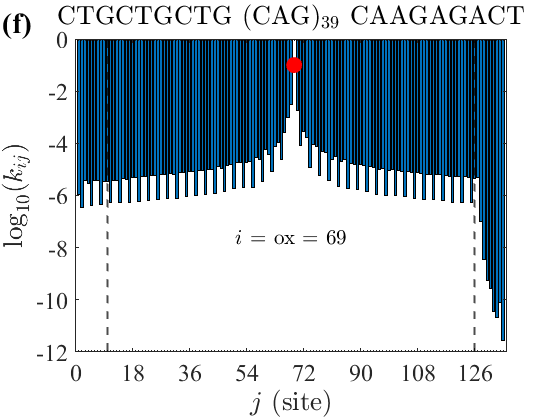}\\
\caption{Kennedy’s disease. Mean transfer rates for: $n=3$ ($N=27$) [left column], and $n=39$ ($N=135$) [right column]. The site of initial oxidation is denoted by the red filled circle. The vertical dashed lines indicate the boundaries between the central periodic region and the primers.
First row: Initial oxidation at the \texttt{C-G} base pair of the central trimer. 
Second row: Initial oxidation at the \texttt{A-T} base pair of the central trimer.
Third row:  Initial oxidation at the \texttt{G-C} base pair of the central trimer.}
\label{fig:KennedyTDkij}
\end{figure}

\begin{figure}[h]
\centering
\includegraphics[width=0.45\textwidth]{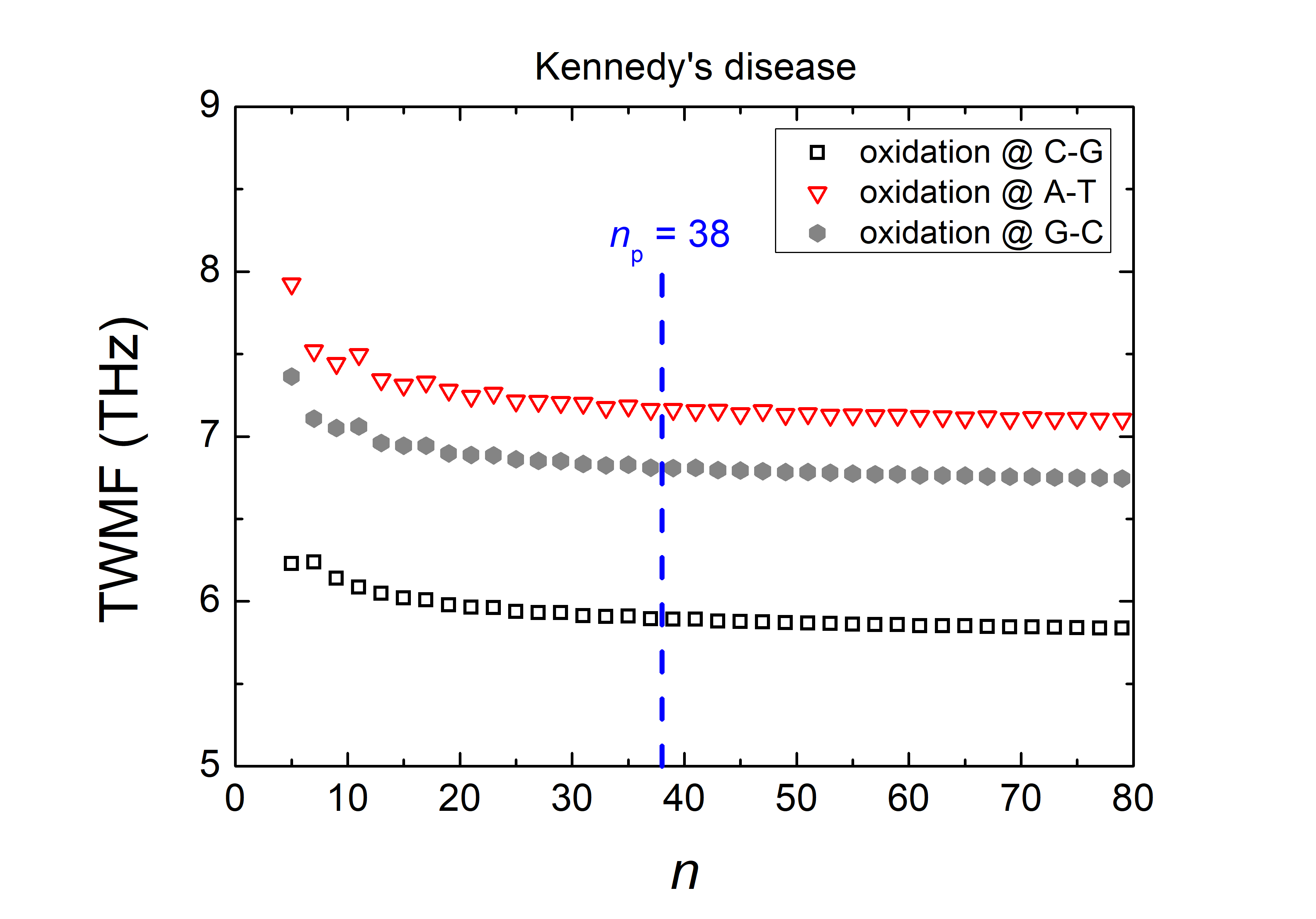}
\caption{Kennedy's disease. Total Weighted Mean Frequency (TWMF), as a function of the number of repeats, $n$. Initial oxidation is performed at the \texttt{C-G} or \texttt{A-T} or \texttt{G-C} base pair of the central \texttt{CAG} trimer.}
\label{fig:KennedyTWMF}
\end{figure}

\newpage


\begin{figure}[h]
\centering
\hspace{-0.5cm}
\includegraphics[width=0.24\textwidth]{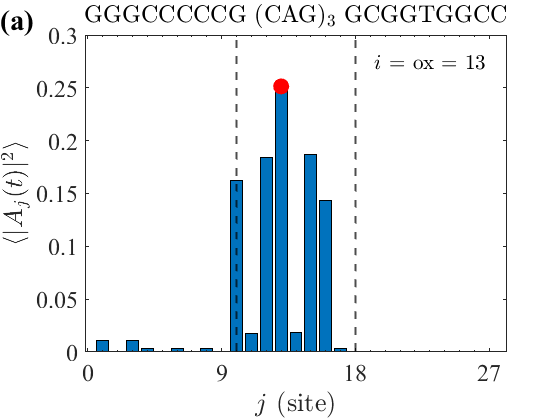}
\includegraphics[width=0.24\textwidth]{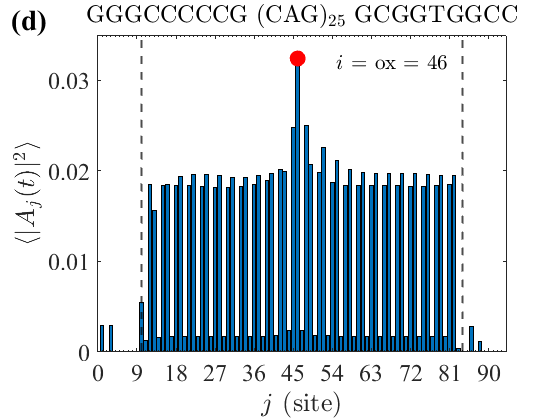}\\
\hspace{-0.5cm}
\includegraphics[width=0.24\textwidth]{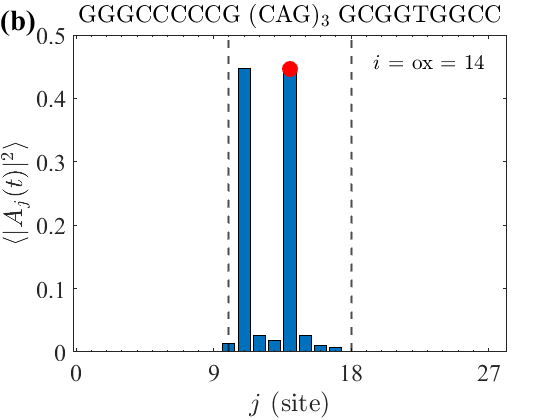}
\includegraphics[width=0.24\textwidth]{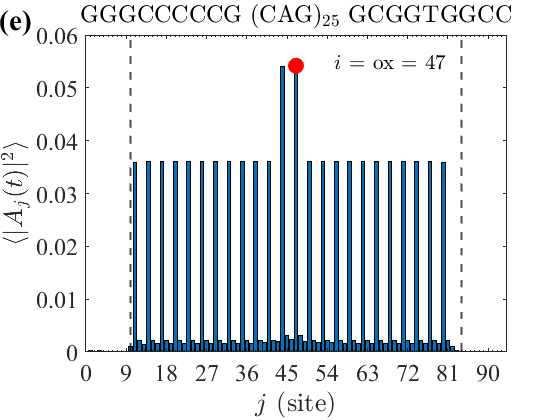}\\
\hspace{-0.5cm}
\includegraphics[width=0.24\textwidth]{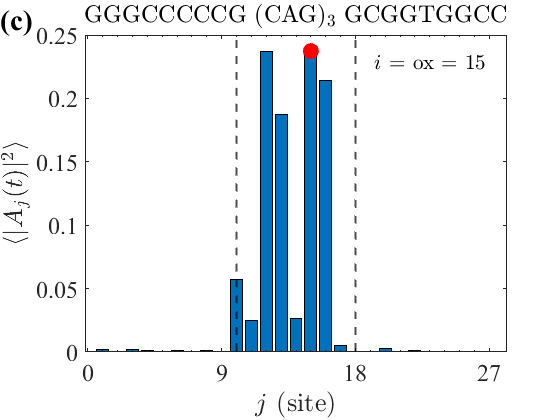}
\includegraphics[width=0.24\textwidth]{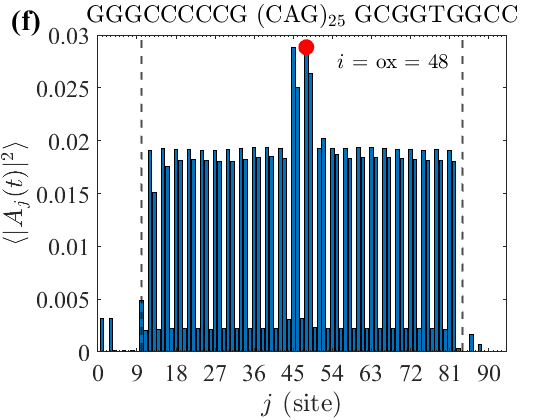}\\
\caption{Spinocerebellar Ataxia 6. Mean over time probabilities for: $n=3$ ($N=27$) [left column], and $n=25$ ($N=93$) [right column].
The site of initial oxidation is denoted by the red filled circle. The vertical dashed lines indicate the boundaries between the central periodic region and the primers.
First row: Initial oxidation at the \texttt{C-G} base pair of the central trimer. 
Second row: Initial oxidation at the \texttt{A-T} base pair of the central trimer.
Third row:  Initial oxidation at the \texttt{G-C} base pair of the central trimer.}
\label{fig:Ataxia6TDmeanprob}
\end{figure}

\begin{figure}[h]
\centering
\includegraphics[width=0.45\textwidth]{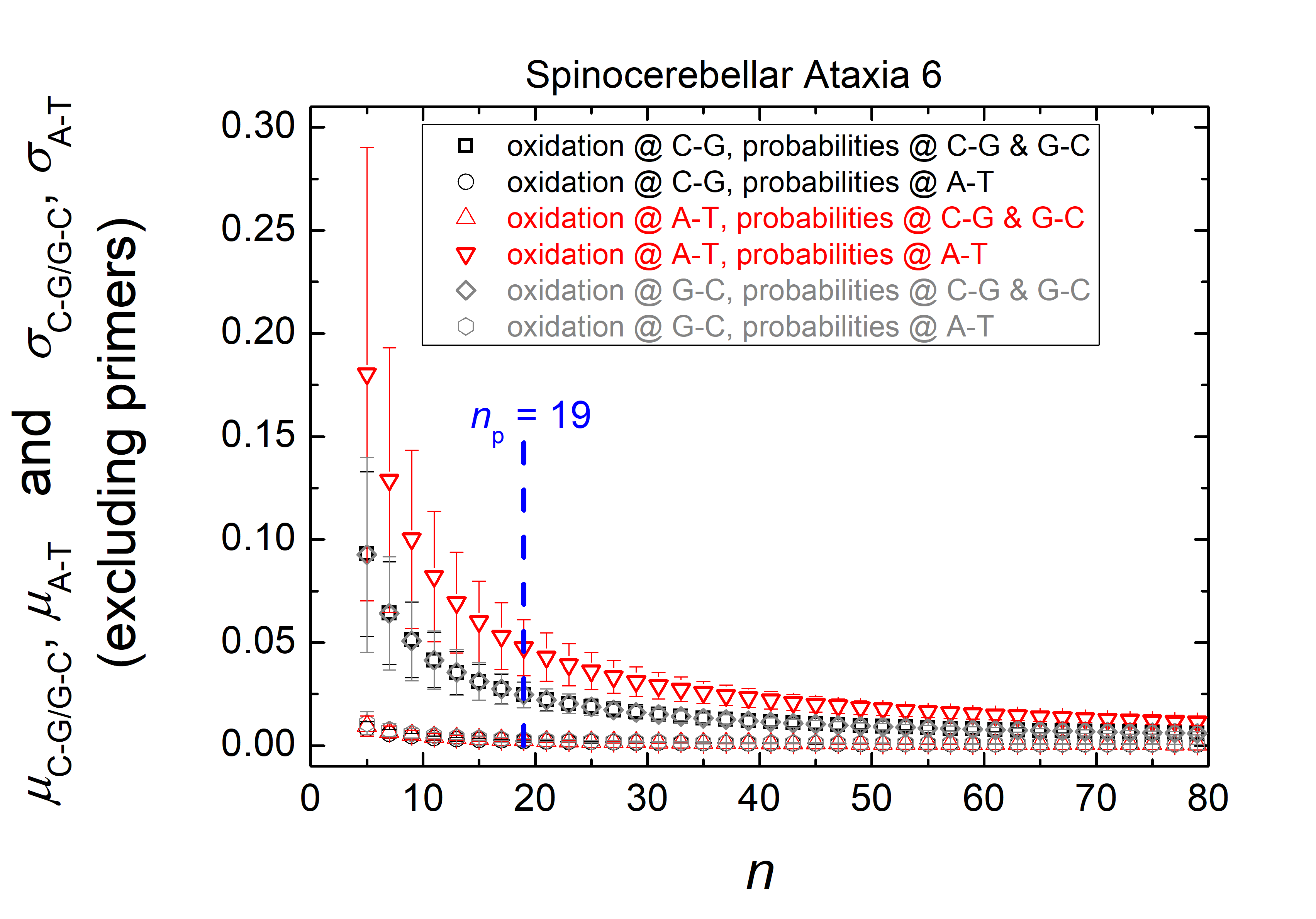}
\caption{Spinocerebellar Ataxia 6. After excluding primers, which act as barriers: Spatial average of mean over time probabilities at \texttt{C-G} or \texttt{G-C} sites, $\mu_\texttt{C-G/G-C} := \overline{\langle |C_j(t)|^2\rangle_\texttt{C-G or G-C}}$, and at \texttt{A-T} sites $\mu_\texttt{A-T} := \overline{\langle |C_\mu(t)|^2\rangle_\texttt{A-T}}$ as well as their corresponding standard deviations, $\sigma_\texttt{C-G/G-C}$ and $\sigma_\texttt{A-T}$, respectively, 
as functions of the number of repeats, $n$. Initial oxidation is performed at the \texttt{C-G} or \texttt{A-T} or \texttt{G-C} base pair of the central (CAG) trimer.}
\label{fig:Ataxia6MeanOfMeans}
\end{figure}

\begin{figure}[h]
\centering
\hspace{-0.5cm}
\includegraphics[width=0.24\textwidth]{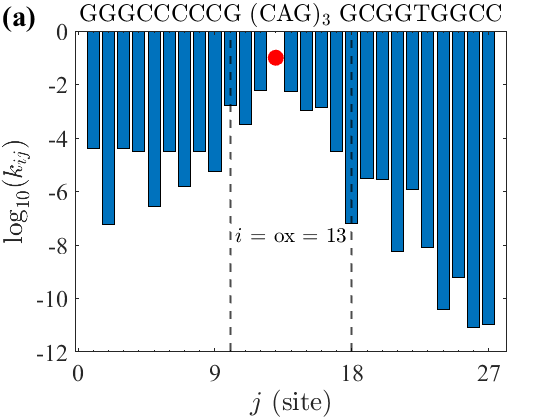}
\includegraphics[width=0.24\textwidth]{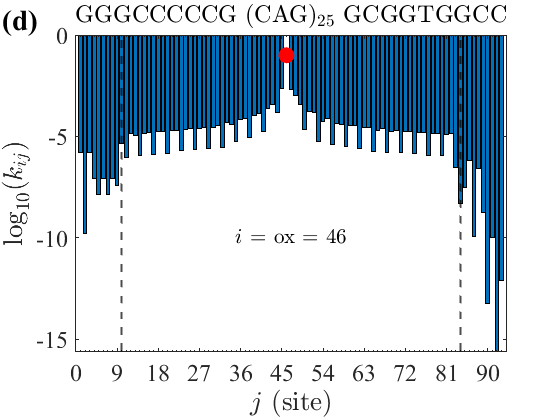}\\
\hspace{-0.5cm}
\includegraphics[width=0.24\textwidth]{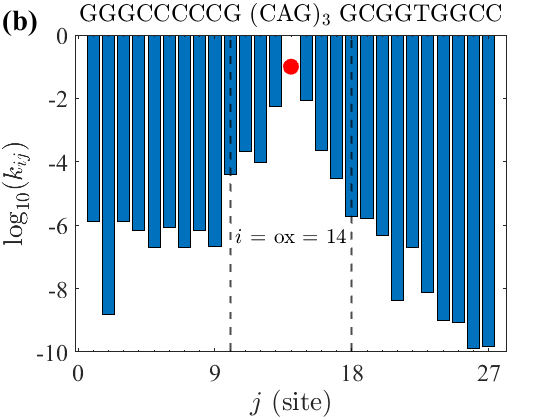}
\includegraphics[width=0.24\textwidth]{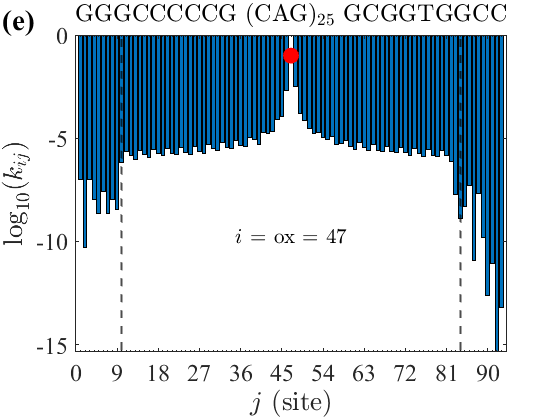}\\
\hspace{-0.5cm}
\includegraphics[width=0.24\textwidth]{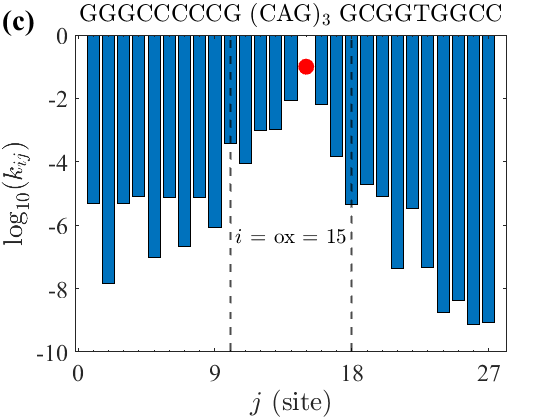}
\includegraphics[width=0.24\textwidth]{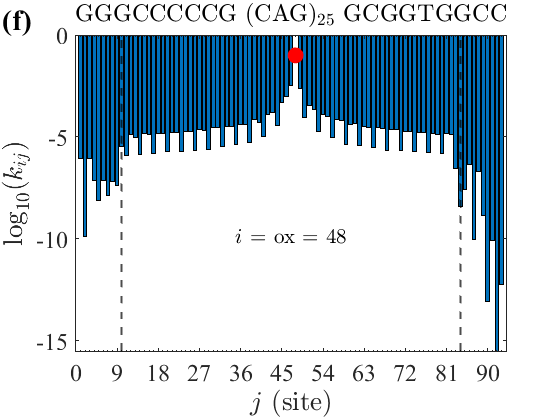}
\caption{Spinocerebellar Ataxia 6. Mean transfer rates for: $n=3$ ($N=27$) [left column], $n=25$ ($N=93$) [right column]. The site of initial oxidation is denoted by the red filled circle. The vertical dashed lines indicate the boundaries between the central periodic region and the primers.
First row: Initial oxidation at the \texttt{C-G} base pair of the central trimer. 
Second row: Initial oxidation at the \texttt{A-T} base pair of the central trimer.
Third row:  Initial oxidation at the \texttt{G-C} base pair of the central trimer.}
\label{fig:Ataxia6TDkij}
\end{figure}

\begin{figure}[h]
\centering
\includegraphics[width=0.45\textwidth]{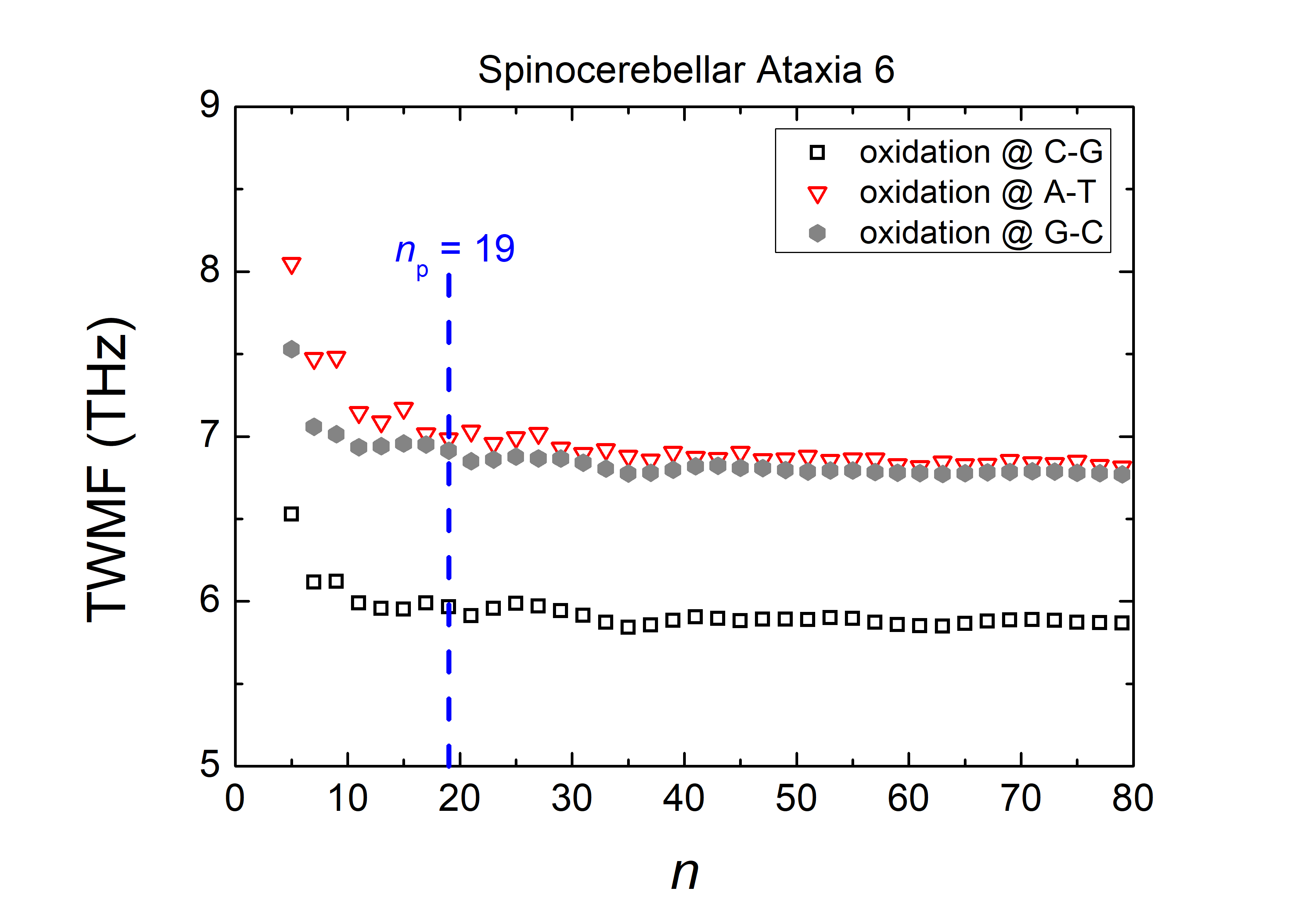}
\caption{Spinocerebellar Ataxia 6. Total Weighted Mean Frequency (TWMF), as a function of the number of repeats, $n$. Initial oxidation is performed at the \texttt{C-G} or \texttt{A-T} or \texttt{G-C} base pair of the central \texttt{CAG} trimer.}
\label{fig:Ataxia6TWMF}
\end{figure}

\newpage


\begin{figure}[!h]
\centering
\hspace{-0.5cm}
\includegraphics[width=0.24\textwidth]{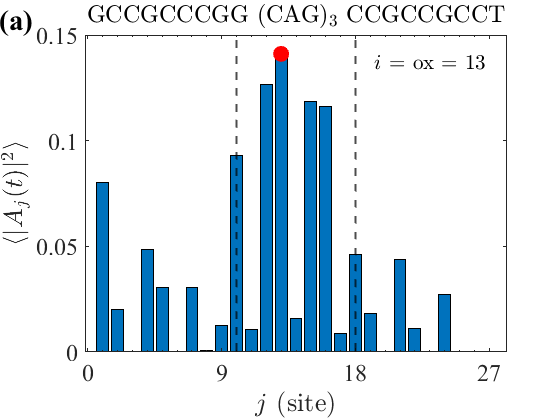}
\includegraphics[width=0.24\textwidth]{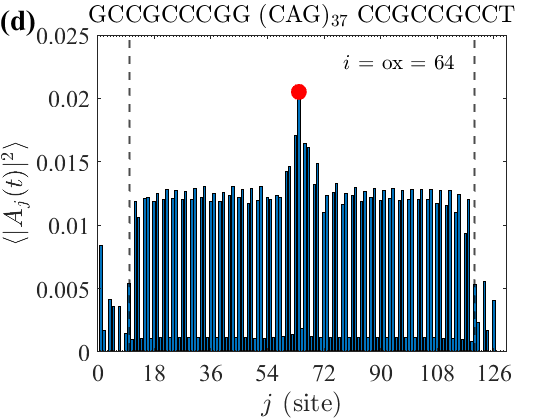}\\
\hspace{-0.5cm}
\includegraphics[width=0.24\textwidth]{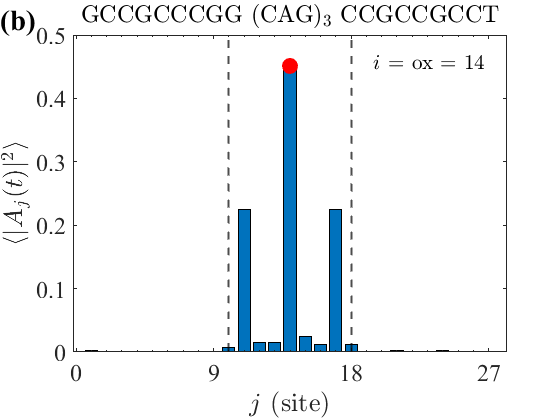}
\includegraphics[width=0.24\textwidth]{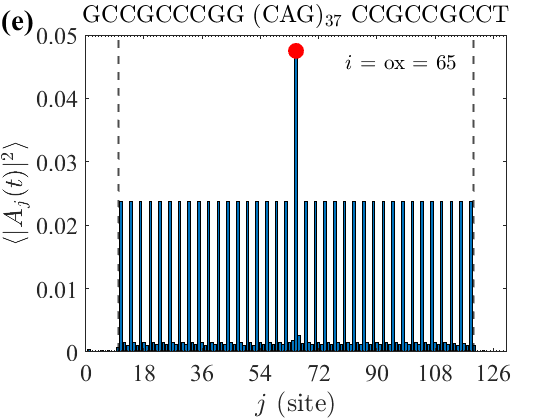}\\
\hspace{-0.5cm}
\includegraphics[width=0.24\textwidth]{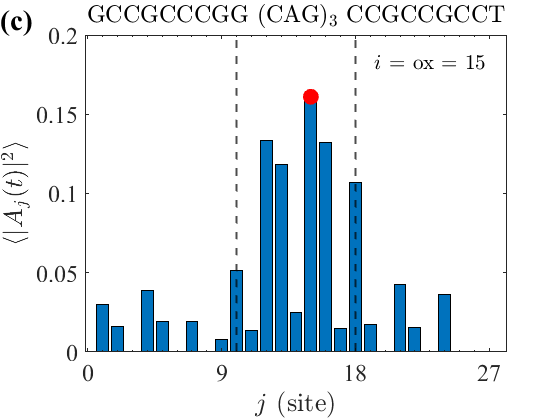}
\includegraphics[width=0.24\textwidth]{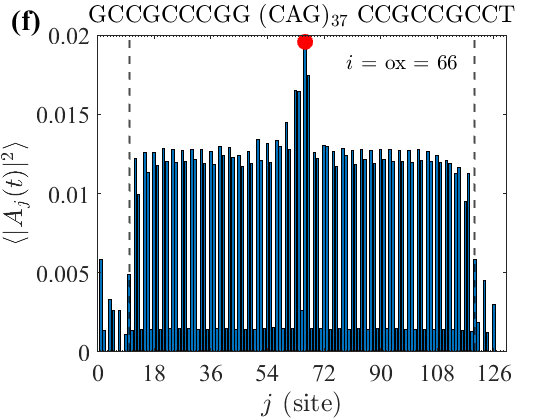}
\caption{Spinocerebellar Ataxia 7. Mean over time probabilities for: $n=3$ ($N=27$) [left column], and $n=37$ ($N=129$) [right column]. 
The site of initial oxidation is denoted by the red filled circle. The vertical dashed lines indicate the boundaries between the central periodic region and the primers.
First row: Initial oxidation at the \texttt{C-G} base pair of the central trimer. 
Second row: Initial oxidation at the \texttt{A-T} base pair of the central trimer.
Third row:  Initial oxidation at the \texttt{G-C} base pair of the central trimer.}
\label{fig:Ataxia7TDmeanprob}
\end{figure}

\begin{figure}[!t]
\centering
\includegraphics[width=0.45\textwidth]{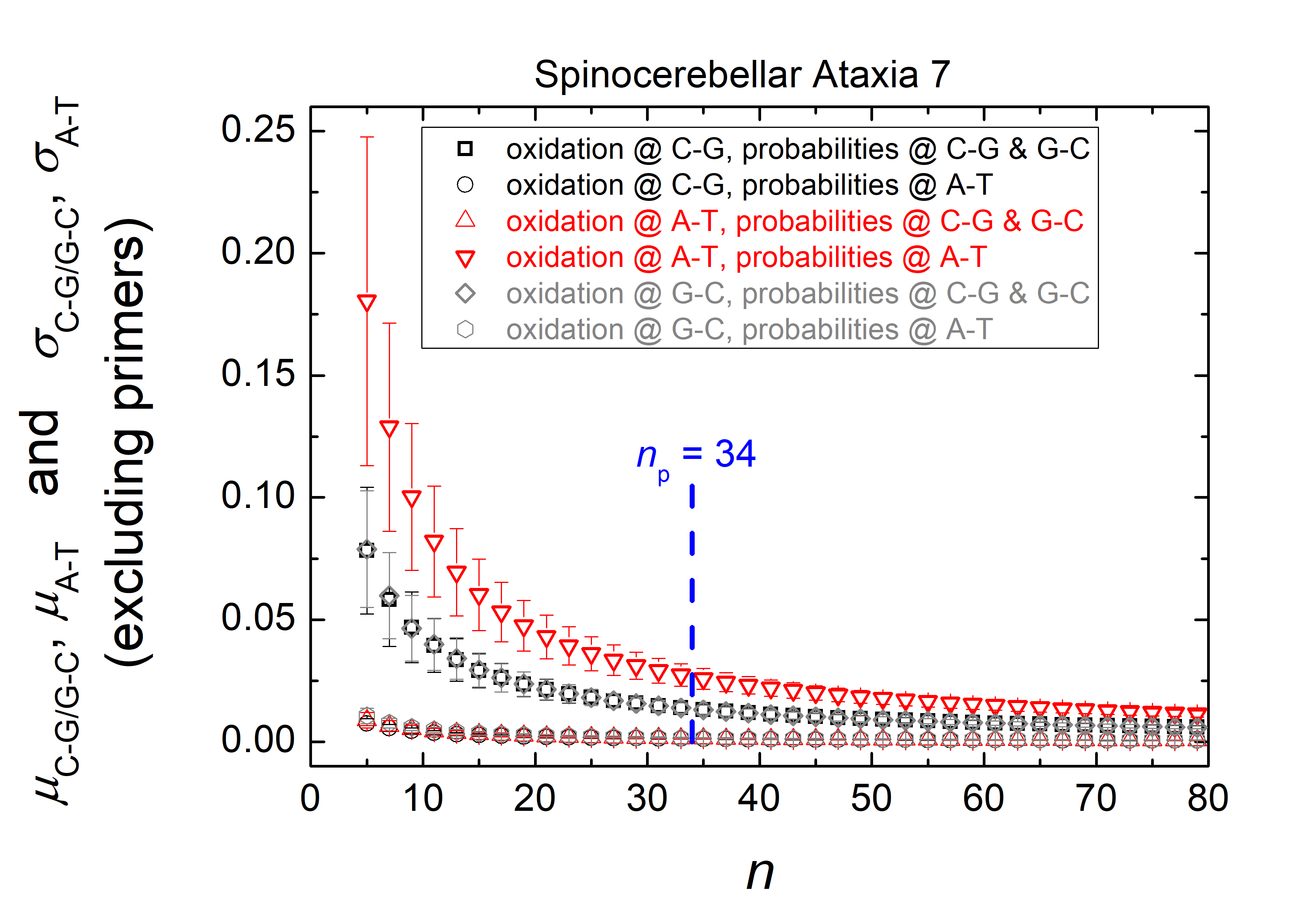}
\caption{Spinocerebellar Ataxia 7. After excluding primers, which act as barriers: Spatial average of mean over time probabilities at \texttt{C-G} or \texttt{G-C} sites, $\mu_\texttt{C-G/G-C} := \overline{\langle |C_j(t)|^2\rangle_\texttt{C-G or G-C}}$, and at \texttt{A-T} sites $\mu_\texttt{A-T} := \overline{\langle |C_\mu(t)|^2\rangle_\texttt{A-T}}$ as well as their corresponding standard deviations, $\sigma_\texttt{C-G/G-C}$ and $\sigma_\texttt{A-T}$, respectively, as functions of the number of repeats, $n$. Initial oxidation is performed at the \texttt{C-G} or \texttt{A-T} or \texttt{G-C} base pair of the central (CAG) trimer.}
\label{fig:Ataxia7MeanOfMeans}
\end{figure}

\begin{figure}[!h]
\centering
\hspace{-0.5cm}
\includegraphics[width=0.24\textwidth]{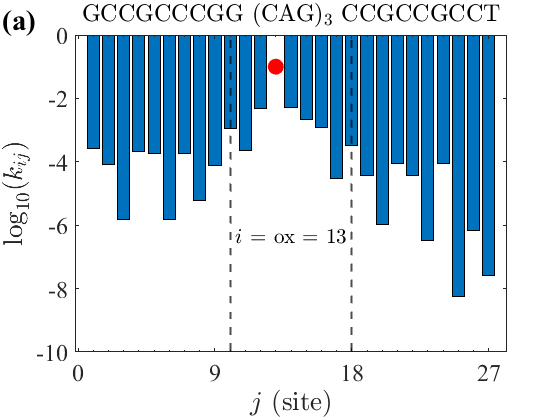}
\includegraphics[width=0.24\textwidth]{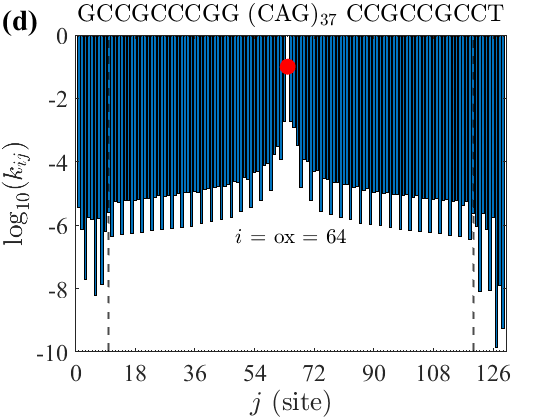}\\
\hspace{-0.5cm}
\includegraphics[width=0.24\textwidth]{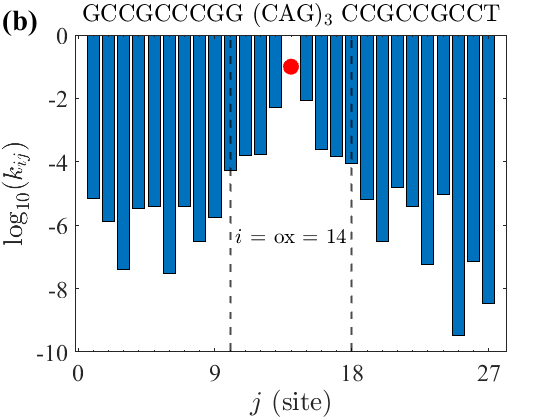}
\includegraphics[width=0.24\textwidth]{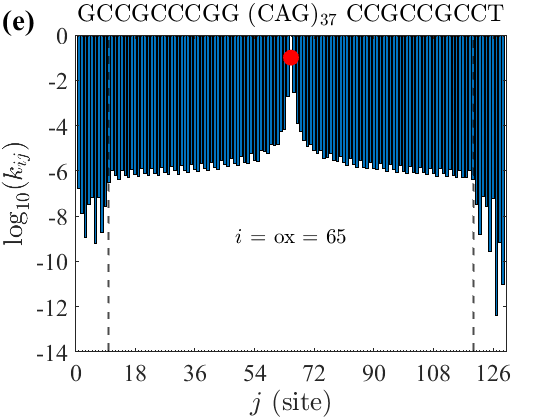}\\
\hspace{-0.5cm}
\includegraphics[width=0.24\textwidth]{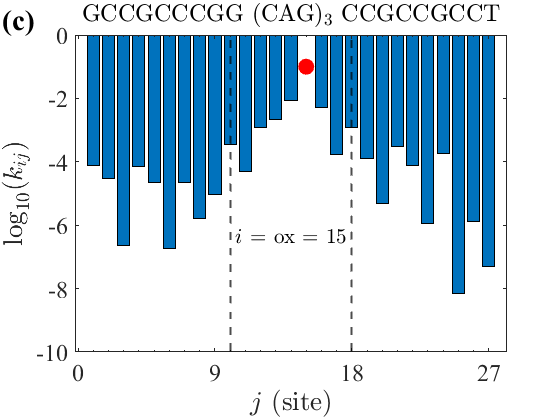}
\includegraphics[width=0.24\textwidth]{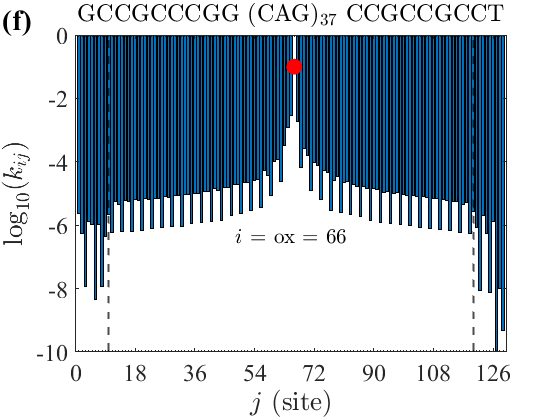}\\
\caption{Spinocerebellar Ataxia 7. Mean transfer rates for: $n=3$ ($N=27$) [left column], and $n=37$ ($N=129$) [right column].The site of initial oxidation is denoted by the red filled circle. The vertical dashed lines indicate the boundaries between the central periodic region and the primers.
First row: Initial oxidation at the \texttt{C-G} base pair of the central trimer. 
Second row: Initial oxidation at the \texttt{A-T} base pair of the central trimer.
Third row:  Initial oxidation at the \texttt{G-C} base pair of the central trimer.}
\label{fig:Ataxia7TDkij}
\end{figure}

\begin{figure}[!t]
\centering
\includegraphics[width=0.45\textwidth]{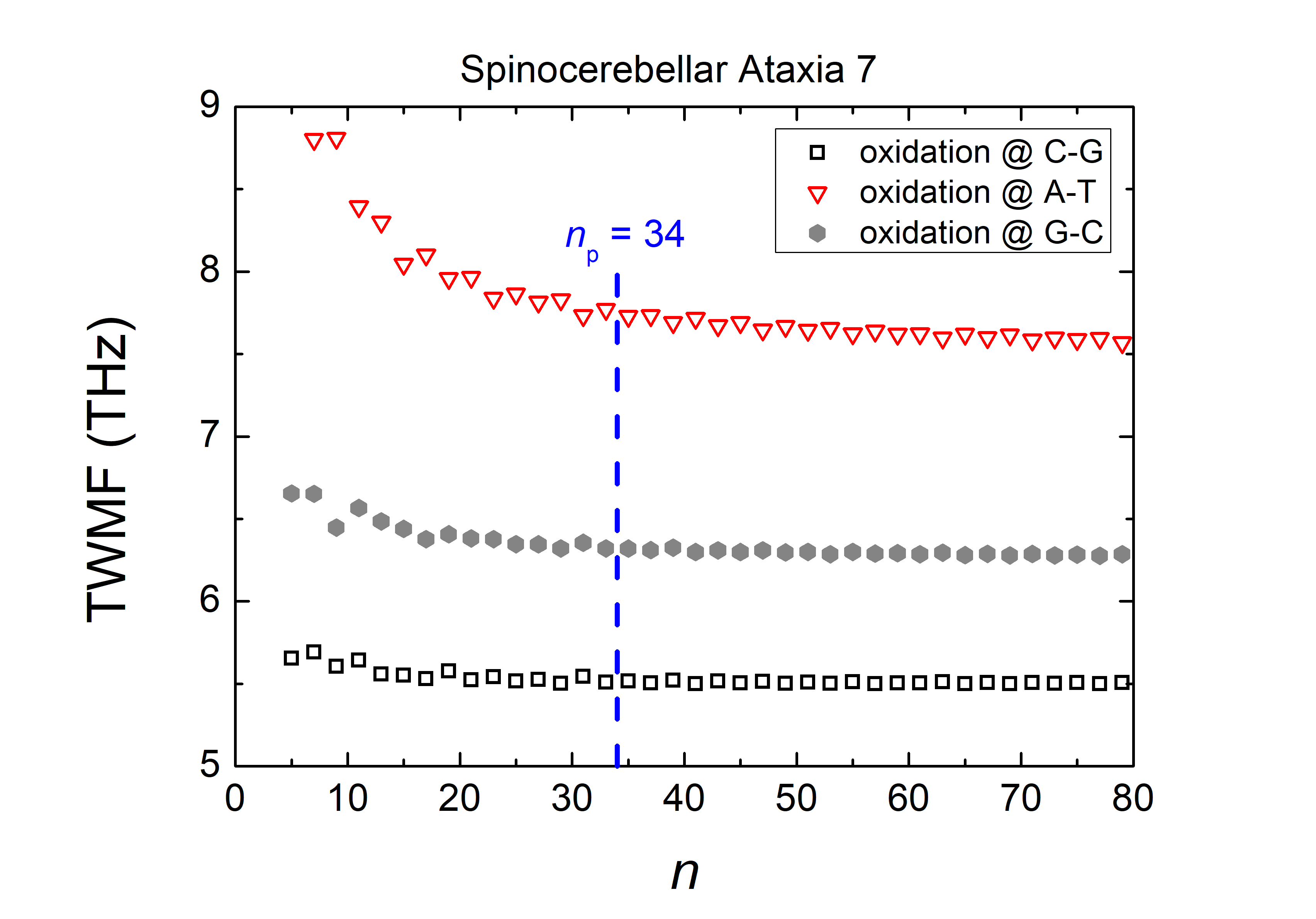}
\caption{Spinocerebellar Ataxia 7. Total Weighted Mean Frequency (TWMF), as a function of the number of repeats, $n$. Initial oxidation is performed at the \texttt{C-G} or \texttt{A-T} or \texttt{G-C} base pair of the central \texttt{CAG} trimer.}
\label{fig:Ataxia7TWMF}
\end{figure}

\clearpage


\end{document}